\documentclass[twocolumn,prb,aps,superscriptaddress,fleqn,eqsecnum,amsmath,amssymb,floats,secnumarabic,floatfix]{revtex4}
\usepackage[latin9]{inputenc}
\usepackage{amstext}
\usepackage{amssymb}
\usepackage{graphicx}

\makeatletter
\@ifundefined{textcolor}{}
{%
 \definecolor{BLACK}{gray}{0}
 \definecolor{WHITE}{gray}{1}
 \definecolor{RED}{rgb}{1,0,0}
 \definecolor{GREEN}{rgb}{0,1,0}
 \definecolor{BLUE}{rgb}{0,0,1}
 \definecolor{CYAN}{cmyk}{1,0,0,0}
 \definecolor{MAGENTA}{cmyk}{0,1,0,0}
 \definecolor{YELLOW}{cmyk}{0,0,1,0}
}

\usepackage{amsfonts}
\usepackage{dcolumn}
\usepackage{array}
\usepackage{bm}

\makeatother

\begin{document}

\title{Emery vs. Hubbard model for cuprate superconductors:\\
a Composite Operator Method study}

\author{Adolfo Avella}

\affiliation{Dipartimento di Fisica ``E.R. Caianiello'', Università degli Studi
di Salerno, I-84084 Fisciano (SA), Italy}

\affiliation{Istituto Internazionale per gli Alti Studi Scientifici (IIASS), I-84019
Vietri sul Mare (SA), Italy}

\affiliation{Unità CNISM di Salerno, Università degli Studi di Salerno, I-84084
Fisciano (SA), Italy}

\affiliation{CNR-SPIN, UoS di Salerno, I-84084 Fisciano (SA), Italy}

\author{Ferdinando Mancini}

\affiliation{Dipartimento di Fisica ``E.R. Caianiello'', Università degli Studi
di Salerno, I-84084 Fisciano (SA), Italy}

\affiliation{Istituto Internazionale per gli Alti Studi Scientifici (IIASS), I-84019
Vietri sul Mare (SA), Italy}

\affiliation{Unità CNISM di Salerno, Università degli Studi di Salerno, I-84084
Fisciano (SA), Italy}

\author{Francesco Paolo Mancini}

\affiliation{Dipartimento di Fisica ``E.R. Caianiello'', Università degli Studi
di Salerno, I-84084 Fisciano (SA), Italy}

\author{Evgeny Plekhanov}

\affiliation{CNR-SPIN, UoS dell'Aquila, I-67010 Coppito (AQ), Italy}
\begin{abstract}
Within the Composite Operator Method (COM), we report the solution
of the Emery model (also known as $p$-$d$ or three band model),
which is relevant for the cuprate high-$T_{c}$ superconductors. We
also discuss the relevance of the often-neglected direct oxygen-oxygen
hopping for a more accurate, sometimes unique, description of this
class of materials. The benchmark of the solution is performed by
comparing our results with the available quantum Monte Carlo ones.
Both single-particle and thermodynamic properties of the model are
studied in detail. Our solution features a metal-insulator transition
at half filling. The resulting metal-insulator phase diagram agrees
qualitatively very well with the one obtained within Dynamical Mean-Field
Theory. We discuss the type of transition (Mott-Hubbard (MH) or charge-transfer
(CT)) for the microscopic (ab-initio) parameter range relevant for
cuprates getting, as expected a CT type. The emerging single-particle
scenario clearly suggests a very close relation between the relevant
sub-bands of the three- (Emery) and the single- band (Hubbard) models,
thus providing an independent and non-perturbative proof of the validity
of the mapping between the two models for the model parameters optimal
to describe cuprates. Such a result confirms the emergence of the
Zhang-Rice scenario, which has been recently questioned. We also report
the behavior of the specific heat and of the entropy as functions
of the temperature on varying the model parameters as these quantities,
more than any other, depend on and, consequently, reveal the most
relevant energy scales of the system.
\end{abstract}
\maketitle

\section{Introduction}

In the last decades, the quest for higher and higher-$T_{c}$ superconductors
and the challenge posed by their microscopical understanding have
been, with no doubts, the hottest topics in solid state and condensed
matter physics~\cite{Dagotto_94,Damascelli_03,Lee_06,Armitage_10}.
The most famous class in the large, and constantly growing, family
of high-temperature superconductors is, by far, the one of cuprate-based
superconductors. It is widely accepted, although sometimes questioned,
that the essential physics in cuprate superconductors takes place
in the $CuO_{2}$ planes~\cite{Anderson_87}. The most general model
describing the interplay between the high level of hybridization of
copper $3d_{x^{2}-y^{2}}$ and oxygen $2p_{\sigma}$ orbitals on one
side and the strong on-site repulsion on copper sites on the other
side, has been proposed by Emery~\cite{Emery_87,Emery_88}, Varma~\cite{Varma_87}
and Loktev~\cite{Gaididei_88,Loktev_96}, and is know as Emery, three-band
or $p$-$d$~model. Zhang and Rice~\cite{Zhang_88} were the first
to argue that upon doping this system with holes, these latter will
occupy the oxygen $2p_{\sigma}$ orbitals, because of the substantial
Coulomb repulsion at the copper sites, and form singlets (the Zhang-Rice
(ZR) singlet (ZRS)) with the holes localized at the copper sites in
$3d_{x^{2}-y^{2}}$ orbitals within the undoped insulating antiferromagnetic
system. For not-so-large values of doping, such singlets are expected
to be the relevant quasi-particles of the system and to embody its
low-energy physics~\cite{Barabanov_97,Kuzian_98}. Such a scenario
motivated several mappings of this (three-band) model to an effective
single-band one (\textit{e.g.}~more or less extended Hubbard~\cite{Zhang_89a,Bacci_91,Schuttler_92,Hayn_93,Batista_93,Simon_93,Feiner_96}
or $t$-$J$~\cite{Emery_88,Ogata_88,Jefferson_92,Yushankhai_97}
models). These mappings return highly non-trivial models, still properly
describing the most relevant and essential physics of cuprates, but
rather easier to be tackled than the original Emery model because
of the greatly reduced number of degrees of freedom (part of the oxygen
degrees of freedom are integrated out and the remaining ones are merged
with the copper ones to give birth to the ZRS as unique effective
orbital per plaquette -- no more oxygen, no more copper). Unfortunately,
the momentum, energy and doping ranges of validity of these mappings
are not known \textit{a priori}, but just roughly guessable, and the
mere existence of the ZRS in overdoped regime is currently under debate~\cite{Peets_09,Phillips_10,Peets_10}.
According to this, in the present article, we will consider the Emery
(three-band) Hamiltonian, and not the Hubbard (single-band) one, because:
(i) we wish to report on its solution in detail, on the interesting
results we found within the COM framework and, in particular, on the
necessity to take into account the direct oxygen-oxygen hopping, often
neglected with no justification, in order to properly describe actual
cuprates, (ii) we wish to provide an independent and non-perturbative
check of the overall validity of the Emery--Hubbard mappings, to determine,
wherever possible by induction, their (parameter) ranges of validity,
and to witness the emergence, if so, of the ZR scenario.

The Emery model represents a true and intriguing challenge for condensed
matter theorists since its proposition: in the absence of an exact
analytical solution, various approximate ones exist. Among the first
approximate, analytical methods applied to the Emery model, we have
to mention the self-energy perturbation theory~\cite{Hotta_94},
the generalized random phase approximation~\cite{Takimoto_97,Takimoto_98}
and the fluctuation exchange (FLEX) approximation~\cite{Koikegami_00,Kobayashi_98}.
As optimal practice, any approximate, analytical method should first
aim at reproducing the results of the mutually complementary numerical
techniques, in the range of their validity, for the very same model.
This allows: (i) to check the true capabilities of the analytical
method to catch some of the physics contained in the chosen model,
but even more important (ii) to discriminate between the failures
of the method and the failures of the model to describe the real material
under analysis up to individuate false agreements driven by the analytical
method artifacts. Numerically, the most interesting region of the
parameter space of the Emery model cannot be accessed directly: quantum
Monte Carlo (qMC) methods suffer from the infamous sign problem and
they are confined either in the weak-coupling and/or in the high-temperature
regimes, or in the small cluster limit~\cite{Dopf_90,Scalettar_91,Dopf_92,Dopf_92a,Kuroki_96}.
In order to circumvent the sign problem, a number of approximate Monte
Carlo techniques have been applied to the Emery model,~\textit{e.g.}~the
variational Monte Carlo method~\cite{Yanagisawa_01,Yanagisawa_09}
and the Constraint Path Monte Carlo technique~\cite{Guerrero_98,Huang_01a}.
Dynamical Mean Field Theory (DMFT) methods~\cite{Zolfl_00,Ono_01,Kent_08,Weber_08,deMedici_09,Wang_10}
are capable to solve the model exactly in the limit of infinite dimensions,
while in the case of finite dimensions, the spatial dependence of
the correlation functions appears to be oversimplified. An extension
to DMFT, called Cluster Perturbation Theory~\cite{Gros_93,Dahnken_02},
has also been applied to the Emery model (for a review on quantum
cluster theories see~\cite{Maier_05,Avella_12} and references therein).
Finally, the Density Matrix Renormalization Group (DMRG) providing
an excellent solution for almost every short-range quantum Hamiltonian
in one spatial dimension (1D), becomes prohibitive in 2D, allowing
to simulate with high precision only a few unit cells at the expense
of high computational effort~\cite{Nishimoto_09}. Nevertheless,
some facts have been established with the aid of complementary methods.
It appears that the Emery model undergoes a metal-insulator transition
in specific regions of its parameter space. In particular, one can
distinguish two insulating regimes relevant to transition metal oxides:
the charge-transfer regime and the Mott-Hubbard one~\cite{Zaanen_85}.
On the basis of the current ab-initio estimates for the Hamiltonian
parameters of the Emery model relevant for cuprates~\cite{McMahan_88,Hybertsen_89,Eskes_89,Eskes_90,Feiner_96},
it is now widely accepted that these materials should belong to the
charge-transfer class. On the contrary, a definite answer, from numerical
and non-perturbative analytical methods, about the emergence of long-range
superconductivity in some regions of the parameter space is still
missing or at least highly controversial. For this class of models,
finite pairing correlations are not so difficult to find within qMC,
but it is still quite unclear whether they are long- or short- ranged~\cite{Kuroki_96,Guerrero_98}.

Coming back to the few analytical methods capable to uncover the complex
and unconventional physics hiding behind the deceptive simplicity
of the Emery model and, in general, to properly and effectively analyze
strongly correlated systems from a non-perturbative perspective, the
composite operator method (COM)~\cite{Theory,COM_in_book} is our
method of choice - we first formulated, and continue developing it
- and, accordingly, we will systematically use it in this manuscript.
The COM framework is based on two main ideas: (i) use of propagators
of relevant composite operators as building blocks for any subsequent
approximate calculations; (ii) use of algebra constraints to fix the
representation of the relevant propagators in order to properly preserve
algebraic and symmetry properties; these constraints will also determine
the unknown parameters appearing in the formulation due to the non-canonical
algebra satisfied by the composite operators. In the last fifteen
years, COM has been successfully applied to several models and materials:
Hubbard~\cite{Hub,Avella_03c,Krivenko_04,Odashima_05,Avella_12b},
$p$-$d$~\cite{Fiorentino_01,p-d}, $t$-$J$~\cite{Avella_02a},
$t$-$t'$-$U$~\cite{ttU}, extended Hubbard ($t$-$U$-$V$)~\cite{tUV},
Kondo~\cite{Villani_00}, Anderson~\cite{Anderson}, two-orbital
Hubbard~\cite{2orb,Plekhanov_11}, Ising~\cite{Ising}, $J_{1}-J_{2}$~\cite{Bak_02a,J1J2,Avella_08a},
Hubbard-Kondo~\cite{Avella_06a}, Cuprates~\cite{Cuprates-NCA,Avella_07,Avella_07a,Avella_08,Avella_09},
etc. COM recipe uses two main ingredients~\cite{Theory,COM_in_book}:
\emph{composite} operators and \emph{algebra} constraints. Composite
operators are products of electronic operators and describe the new
elementary excitations appearing in the system owing to strong correlations.
According to the system under analysis \cite{Theory,COM_in_book},
one has to choose a set of composite operators as operatorial basis
and rewrite the electronic operators and the electronic Green's function
in terms of this basis. Algebra constraints are relations among correlation
functions dictated by the non-canonical operatorial algebra closed
by the chosen operatorial basis~\cite{Theory,COM_in_book}. Other
ways to obtain algebra constraints rely on the symmetries enjoined
by the Hamiltonian under study, the Ward-Takahashi identities, the
hydrodynamics, etc~\cite{Theory,COM_in_book}. Algebra constraints
are used to compute unknown correlation functions appearing in the
calculations. Interactions among the elements of the chosen operatorial
basis are described by the residual self-energy, that is, the propagator
of the residual term of the current after this latter has been projected
on the chosen operatorial basis~\cite{Theory,COM_in_book}. According
to the physical properties under analysis and the range of temperatures,
dopings, and interactions to be explored, one has to choose an approximation
to compute the residual self-energy.

\begin{figure}
\includegraphics[angle=270,width=8cm]{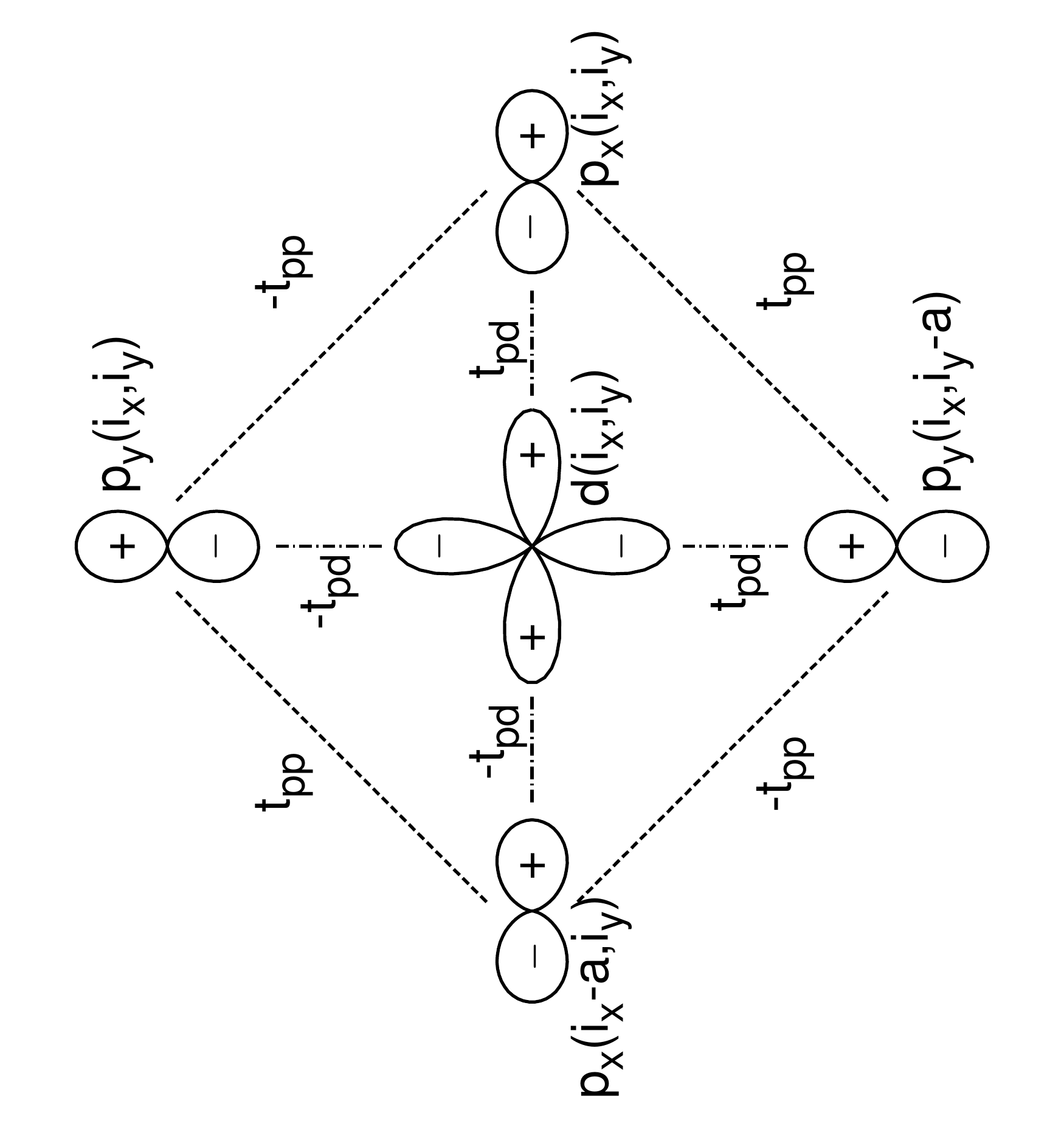} \caption{Orbital configuration of an elementary plaquette centered at a generic
Cu-site of coordinates $\textbf{i}=(i_{x},i_{y})$ and covering the
CuO$_{2}$ plane.}

\label{fig1} 
\end{figure}

It has been shown in~\cite{Fiorentino_01} that the minimal number
of composite operators in the operatorial basis, which is necessary
in order to catch the essential physics of the Emery model, is four.
The first three of them are: the two Hubbard operators, named $\xi$
and $\eta$ within COM, describing the $d$ electrons of copper and
the operator $p$, describing the bonding component of the $p_{\alpha}$
electrons of oxygen. The fourth field, named $p_{s}$, describes the
$p$-electronic excitations dressed by the nearest neighbor (NN) $d$-electron
spin fluctuations; its precise definition will be given in the next
section. Such a basis has already been assessed~\cite{Fiorentino_01}
in the case of the Emery model without direct oxygen-oxygen hopping
term. In this manuscript, (i) we add this latter term to the Hamiltonian
under analysis, as this term appears non-negligible in cuprates from
all available ab-initio estimates (see for instance~\cite{McMahan_88,Hybertsen_89,Eskes_89,Eskes_90,Feiner_96,Kent_08}),
(ii) we propose an alternative choice for the fourth field, which
will take into account not only the $d$-electron spin fluctuations,
but also the charge and pair ones, and (iii) we focus our study onto
the single-particle and the thermodynamic properties of the model,
in order to analyze the dependence of these latter on the additional
hopping term and check the validity of the Zhang and Rice scenario.

The plan of the paper follows: in Sec.~\ref{sec:model}, we discuss
in some detail the Emery model and, in particular, the variant we
decided to focus on; in Sec.~\ref{sec:method}, we describe the application
of the COM to the Emery model introducing different possible choices
for the operatorial basis and the self-consistency scheme; in Sec.~\ref{sec:results:cfr},
in order to benchmark the method and the chosen operatorial basis,
we compare our results to qMC ones; in Sec.~\ref{sec:results:mit},
we characterize the metal-insulator transition (MIT) featured by the
model under study, compare our results to DMFT ones, and check the
influence on the MIT of the direct oxygen-oxygen hopping term; in
Sec.~\ref{sec:results:spzr}, we discuss the emergence of the ZRS
as main actor at low-energy in a specific region of the parameter
space and, accordingly, the soundness of the single-band-model mappings
through the analysis of our results for the single-particle properties
of the model (DOS, bands and their orbital character); in Sec~\ref{sec:results:thermo},
we report our results on the peculiar behaviors of the specific heat
and of the entropy of the model as functions of the temperature on
varying the model parameters and on the strict relationship of these
two thermodynamic quantities to the most relevant energy scales of
the system. Finally, Sec.~\ref{sec:conclu} contains a brief summary
of the manuscript and possible perspectives.

\section{Model\label{sec:model}}

Let us take into account only the $d$- and $p$- electrons in the
Cu$_{d{}_{x^{2}-y^{2}}}$ and O$_{p_{x},p_{y}}$ (Wannier) orbitals,
respectively, of a CuO$_{2}$ plane belonging to a generic single-layer
cuprate (i.e. let us consider the apical oxygens, the bilayer splitting,
the reservoir chains and so forth as secondary players). The related
orbital configuration of an elementary plaquette -- one of those covering
the CuO$_{2}$ plane -- centered at the generic coordinates $\textbf{i}=(i_{x},i_{y})$
within the square Bravais lattice of lattice constant $a$ formed
by the Cu-sites is shown in Fig.~\ref{fig1}. Accordingly, the Hamiltonian
of the Emery model reads as follows:\begin{widetext} 
\begin{equation}
\begin{split}H & =\sum\limits _{\textbf{i}}\left\{ \left(\varepsilon_{d}-\mu\right)d^{\dag}(\textbf{i})d(\textbf{i})+\left(\varepsilon_{p}-\mu\right)\left[p_{x}^{\dag}(\textbf{i})p_{x}(\textbf{i})+p_{y}^{\dag}(\textbf{i})p_{y}(\textbf{i})\right]\right\} +U\sum\limits _{\textbf{i}}n_{\uparrow}^{d}(\textbf{i})n_{\downarrow}^{d}(\textbf{i})\\
 & +t_{pd}\sum\limits _{\textbf{i}}\left\{ d^{\dag}(\textbf{i})\left[p_{x}(\textbf{i})-p_{x}(i_{x}-a,i_{y})-p_{y}(\textbf{i})+p_{y}(i_{x},i_{y}-a)\right]+h.c.\right\} \\
 & +t_{pp}\sum\limits _{\textbf{i}}\left\{ p_{x}^{\dag}(\textbf{i})[-p_{y}(\textbf{i})+p_{y}(i_{x},i_{y}-a)+p_{y}(i_{x}+a,i_{y})-p_{y}(i_{x}+a,i_{y}-a)]+h.c.\right\} 
\end{split}
\label{eq1}
\end{equation}
\end{widetext}The creation operators for the electronic states in
the orbital $f=p_{x},p_{y},d$ are denoted as $f^{\dagger}(i)$ in
spinorial notation: $f^{\dag}(i)=\left(f_{\uparrow}^{\dag}(i),f_{\downarrow}^{\dag}(i)\right)$.
In the Heisenberg picture, we have $i=(\textbf{i},t)$. The field
operators $d(i)$ and $p_{x(y)}(i)$ satisfy canonical anti-commutation
relations. $\mu$ is the chemical potential. $\varepsilon_{d}$ and
$\varepsilon_{p}$ are the Cu$_{d{}_{x^{2}-y^{2}}}$ and O$_{p_{x},p_{y}}$
atomic levels, respectively. $t_{pd}$ and $t_{pp}$ stand for the
$p-d$ and $p-p$ hopping integrals, respectively. $n_{\sigma}^{d}(i)=d_{\sigma}^{\dag}(i)d_{\sigma}(i)$
and $n_{x(y),\sigma}^{p}(i)=p_{x(y),\sigma}^{\dag}(i)p_{x(y),\sigma}(i)$
are the density operators for spin $\sigma$ at the site $\textbf{i}$
of the $d$ and $p$ electrons, respectively. $U$ is the on-site
Coulomb repulsion strength among $d$ electrons with opposite spins.
$n^{d}(i)=\sum_{\sigma}n_{\sigma}^{d}$ and $n_{x(y)}^{p}(i)=\sum_{\sigma}n_{x(y),\sigma}^{p}(i)$
are the total density operators at the site $\textbf{i}$ of the $d$
and $p$ electrons, respectively. We decided to neglect the on-site
Coulomb repulsion among $p$ electrons as well as the inter-site Coulomb
interaction between $p$ and $d$ electrons, as suggested by many
ab-initio calculations (see for instance~\cite{McMahan_88,Hybertsen_89})
reporting negligible strengths for them, with respect to the other
energy scales present in the problem.

The $p_{x}(i)$ and $p_{y}(i)$ operators enter the hybridization
term in such a way that we can simplify the Hamiltonian (\ref{eq1})
by eliminating one degree of freedom. Indeed, let us consider the
Bogolyubov transformation~\cite{Tachiki_90}

\[
\begin{split}p(\textbf{k}) & =\frac{i}{\sqrt{1-\alpha(\textbf{k})}}\left[\sin\left(\frac{k_{x}a}{2}\right)p_{x}(\textbf{k})-\sin\left(\frac{k_{y}a}{2}\right)p_{y}(\textbf{k})\right]\\
q(\textbf{k}) & =\frac{i}{\sqrt{1-\alpha(\textbf{k})}}\left[\sin\left(\frac{k_{y}a}{2}\right)p_{x}(\textbf{k})+\sin\left(\frac{k_{x}a}{2}\right)p_{y}(\textbf{k})\right]\\
\\
\\
\\
\end{split}
\]
constructed in terms of the the Fourier transforms of the electronic
$p_{x}(i)$, $p_{y}(i)$ and $d(i)$ original operators

\[
\begin{split}p_{x(y)}(\textbf{i}) & =\sqrt{\frac{\Omega}{(2\pi)^{2}}}\int\limits _{\Omega_{B}}d^{2}\textbf{k}\exp\left(i\textbf{k}\cdot\textbf{i}+\frac{ik_{x(y)}a}{2}\right)p_{x(y)}(\textbf{k})\\
d(\textbf{i}) & =\sqrt{\frac{\Omega}{(2\pi)^{2}}}\int\limits _{\Omega_{B}}d^{2}\textbf{k}\exp(i\textbf{k}\cdot\textbf{i})d(\textbf{k})
\end{split}
\]
where $\Omega=a^{2}$ and $\Omega_{B}$ are the volumes of the units
cell in the direct and reciprocal spaces, respectively, and $\alpha(\textbf{k})=\frac{1}{2}[\cos(k_{x}a)+\cos(k_{y}a)]$
is the Fourier transform of the projection operator $\alpha_{\textbf{i}\textbf{j}}$
onto the nearest-neighbor sites of a square lattice. Under the above
transformation, the hopping terms in the Hamiltonian (\ref{eq1})
take the form

\begin{widetext} 
\begin{equation}
\begin{split}\sum\limits _{\textbf{i}} & d^{\dag}(\textbf{i})[p_{x}(\textbf{i})-p_{x}(i_{x}-a,i_{y})-p_{y}(\textbf{i})+p_{y}(i_{x},i_{y}-a)]+h.c.=2\int\limits _{\Omega_{B}}d^{2}\textbf{k}\gamma(\textbf{k})[d^{\dag}(\textbf{k})p(\textbf{k})+p^{\dag}(\textbf{k})d(\textbf{k})]\\
\sum\limits _{\textbf{i}} & \{p_{x}^{\dag}(\textbf{i})[-p_{y}(\textbf{i})+p_{y}(i_{x},i_{y}-a)+p_{y}(i_{x}+a,i_{y})-p_{y}(i_{x}+a,i_{y}-a)]+h.c.\\
= & 2\int\limits _{\Omega_{B}}d^{2}\textbf{k}\lambda(\textbf{k})[p^{\dag}(\textbf{k})p(\textbf{k})-q^{\dag}(\textbf{k})q(\textbf{k})]+2\int\limits _{\Omega_{B}}d^{2}\textbf{k}\tau(\textbf{k})[p^{\dag}(\textbf{k})q(\textbf{k})+q^{\dag}(\textbf{k})p(\textbf{k})].
\end{split}
\label{eq4}
\end{equation}
\end{widetext}where
\begin{equation}
\begin{split}\gamma(\textbf{k}) & =\sqrt{1-\alpha(\textbf{k})}\\
\lambda(\textbf{k}) & =\frac{1-2\alpha(\textbf{k})+\beta(\textbf{k})}{1-\alpha(\textbf{k})}\\
\beta(\textbf{k}) & =\cos(k_{x}a)\cos(k_{y}a)\\
\tau(\textbf{k}) & =\frac{\sin\left(\frac{k_{x}a}{2}\right)\sin\left(\frac{k_{y}a}{2}\right)}{1-\alpha(\textbf{k})}[\cos(k_{x}a)-\cos(k_{y}a)].
\end{split}
\label{eq5}
\end{equation}

As Eq.~(\ref{eq4}) clearly shows, $p(\textbf{k})$ and $q(\textbf{k})$
play the roles of bonding and anti-bonding components, respectively,
of the $p_{x}$ and $p_{y}$ fields with respect to the $d$ one:
the non-bonding component $q(\textbf{k})$ is not coupled directly
to $d(\textbf{k})$, but only to the bonding component $p(\textbf{k})$.
Accordingly, as we consider the dynamics of the $q(\textbf{k})$ field
only marginally relevant, we will neglect completely the $q(\textbf{k})$
field hereinafter in order to significantly simplify the Emery Hamiltonian~(\ref{eq1}),
which now reads as
\begin{equation}
\begin{split} & H=\sum\limits _{\textbf{i}}\left[(\varepsilon_{d}-\mu)d^{\dag}(\textbf{i})d(\textbf{i})+(\varepsilon_{p}-\mu)p^{\dag}(\textbf{i})p(\textbf{i})\right]\\
 & +2t_{pd}\sum\limits _{\textbf{i}}[p^{\gamma\dag}(\textbf{i})d(\textbf{i})+d^{\dag}(\textbf{i})p^{\gamma}(\textbf{i})]\\
 & +2t_{pp}\sum\limits _{\textbf{i}}p^{\dag}(\textbf{i})p^{\lambda}(\textbf{i})+U\sum\limits _{\textbf{i}}n_{\uparrow}^{d}(\textbf{i})n_{\downarrow}^{d}(\textbf{i})
\end{split}
\label{eq:ham}
\end{equation}
where we introduced the shorthand notation $\Phi^{\kappa}(i)=\sum\limits _{\mathbf{j}}\kappa_{\mathbf{ij}}\Phi(\mathbf{j},t)$
and $\Phi^{\kappa\rho}(i)=\sum\limits _{\mathbf{jl}}\kappa_{\mathbf{il}}\rho_{\mathbf{lj}}\Phi(\mathbf{j},t)$,
which we will use hereafter for any generic operator $\Phi(i)$.

\section{Method\label{sec:method}}

\subsection{Operatorial basis and equations of motion}

We have solved the Hamiltonian (\ref{eq:ham}) by using the Green's
function and the equations of motion formalisms within the COM framework~\cite{Theory,COM_in_book}.
One of the main ingredients of the method is the extremely sound observation
that, in presence of strong electronic interactions, the focus should
be moved from the bare electronic operators, in terms of which any
perturbative calculation is doomed to fail, to new operators (composite
operators). These latter (i) naturally emerge from the dynamics, (ii)
seamlessly embed, since the very beginning, the interactions, and
(iii) make feasible to avoid the search for and use of unlikely small
parameters. According to this, the very first step to be taken regards
the choice of a suitable basis of composite operators. In this manuscript,
we analyze the pros and cons of two possible choices for such a basis,
both dictated by the hierarchy of the equations of motion.

\subsubsection{Basis I}

As first choice, we consider the following multi-component composite
field operator 
\begin{equation}
\begin{split}\psi(i) & =\left(\begin{array}{c}
p(i)\\
\xi(i)\\
\eta(i)\\
p_{s}(i)
\end{array}\right)\end{split}
\label{eq8}
\end{equation}
where $p(i)$ is the bonding component of the $p$ field, $\xi(i)=[1-n^{d}(i)]d(i)$
and $\eta(i)=n^{d}(i)d(i)$ are the Hubbard operators for the $d$
electrons. The fourth field, $p_{s}(i)$, is defined as follows: 
\begin{equation}
p_{s}(i)=\sigma_{k}n_{k}^{d}(i)p^{\gamma}(i)-\frac{3c}{I_{22}}\xi(i)-\frac{3b}{I_{33}}\eta(i),\label{eq9}
\end{equation}
$n_{\mu}^{d}(i)=d^{\dag}(i)\sigma_{\mu}d(i)$ is the charge and spin
number operator of the $d$ electrons, $\sigma_{\mu}=(1,\vec{\sigma})$,
and $\sigma^{\mu}=(-1,\vec{\sigma})$, $\sigma_{k}$ being the Pauli
matrices. The parameters $b$ and $c$ and the quantities $I_{22}$
and $I_{33}$ are defined in Appendix~\ref{app:basisI}. The choice
of the multi-component composite field operator~(\ref{eq8}) is dictated
by the following considerations. The quite strong on-site Coulomb
repulsion $U$ at Cu ions causes the splitting of the $d$ band into
the lower and the upper Hubbard sub-bands. These latter are exactly
described by the $\xi$ and $\eta$ Hubbard operators, which are capable
to distinguish among no-, single- and double- occupancy of a site,
unlike canonical $d$ operator. Such a capability simply puts $\xi$
and $\eta$ in the position to properly take into account the scale
of energy of $U$. The very high degree of covalence between oxygen
and copper orbitals leads to large fluctuations in the energy of $p$
electrons, whose dynamics turns out to be strongly entangled to that
of $d$ electrons. In particular, the electronic $p$ excitations
are strongly affected by the charge, spin and pair $d$ excitations,
see Eq.~(\ref{eq:em1}) below. Accordingly, it is now evident the
fundamental relevance of the field $p_{s}(i)$, describing the $p$
electronic excitations dressed by the nearest-neighbor $d$ electron
spin fluctuations.

Given the Hamiltonian (\ref{eq:ham}), we obtain the following equations
of motion for the basic field $\psi(i)$

\begin{eqnarray*}
\end{eqnarray*}
\begin{eqnarray}
 & i\frac{\partial}{\partial t}\psi(i) & =[\psi(i),H]=J(i)\nonumber \\
 & = & \left(\begin{array}{l}
{(\varepsilon_{p}-\mu)p(i)+2t_{pp}p^{\lambda}(i)+2t_{pd}[\xi^{\gamma}(i)+\eta^{\gamma}(i)]}\\
\\
{2t_{pd}p^{\gamma}(i)+(\varepsilon_{d}-\mu)\xi(i)+2t_{pd}\pi(i)}\\
\\
{(\varepsilon_{d}-\mu+U)\eta(i)-2t_{pd}\pi(i)}\\
\\
\varepsilon_{pp}p^{\gamma}(i)+\varepsilon_{p\xi}\xi(i)+\varepsilon_{p\eta}\eta(i)+(\varepsilon_{p}-\mu)p_{s}(i)\\
+t_{p}\pi(i)+2t_{pd}\kappa_{s}(i)+2t_{pp}\lambda_{s}(i)
\end{array}\right)\nonumber \\
\label{eq:em1}
\end{eqnarray}
where the following higher-order composite fields appear
\begin{equation}
\begin{split}\pi(i) & =\frac{1}{2}\sigma^{\mu}n_{\mu}^{d}(i)p^{\gamma}(i)+\xi(i)p^{\gamma\dag}(i)\eta(i)\\
\lambda_{s}(i) & =\sigma_{k}n_{k}^{d}(i)p^{\gamma\lambda}(i)\\
\kappa_{s}(i) & =\sigma_{k}d^{\dag}(i)\sigma_{k}p^{\gamma}(i)p^{\gamma}(i)-\sigma_{k}p^{\gamma\dag}(i)\sigma_{k}d(i)p^{\gamma}(i)\\
 & +\sigma_{k}n_{k}^{d}(i)d^{\gamma\gamma}(i)
\end{split}
\label{eq11}
\end{equation}
and the following notation is used
\begin{equation}
t_{p}=6t_{pd}\left(\frac{b}{I_{33}}-\frac{c}{I_{22}}\right),\quad\varepsilon_{pp}=-\frac{6t_{pd}c}{I_{22}}
\end{equation}
\begin{equation}
\varepsilon_{p\xi}=\frac{3c}{I_{22}}\left(U-\Delta\right),\quad\varepsilon_{p\eta}=-\frac{3b}{I_{33}}\Delta,
\end{equation}
with $\Delta=U-(\varepsilon_{p}-\varepsilon_{d})$, which is the charge
transfer gap in the electronic representation.

\subsubsection{Basis II}

As second choice, we consider the following multiplet composite field
\begin{equation}
\psi(i)=\left(\begin{array}{c}
p(i)\\
\xi(i)\\
\eta(i)\\
\hat{\pi}(i)
\end{array}\right)\label{eq80}
\end{equation}
The first three fields coincide with those defined above in Eq.~(\ref{eq8}),
while the fourth field $\hat{\pi}(i)$ is defined as follows: 
\begin{equation}
\hat{\pi}(i)=\pi(i)+I_{33}p^{\gamma}(i)-g\left(\frac{\xi(i)}{I_{22}}-\frac{\eta(i)}{I_{33}}\right)\label{eq14}
\end{equation}
where $\pi(i)$ is defined in Eq.~(\ref{eq11}) and the parameter
$g$ is given by $g=c-b$. In addition to the spin fluctuations, the
field $\hat{\pi}(i)$ takes into account charge and pair fluctuations
too.

The equations of motion for this second basic field read as

\begin{widetext} 
\begin{equation}
i\frac{\partial}{\partial t}\psi(i)=J(i)=\left(\begin{array}{l}
{(\varepsilon_{p}-\mu)p(i)+2t_{pp}p^{\lambda}(i)+2t_{pd}[\xi^{\gamma}(i)+\eta^{\gamma}(i)]}\\
\\
{[\varepsilon_{d}-\mu+2t_{pd}I_{22}^{-1}g]\xi(i)+2t_{pd}[I_{22}p^{\gamma}(i)-I_{33}^{-1}g\eta(i)+\hat{\pi}(i)]}\\
\\
{[\varepsilon_{d}-\mu+U+2t_{pd}I_{33}^{-1}g]\eta(i)+2t_{pd}[I_{33}p^{\gamma}(i)-I_{22}^{-1}g\xi(i)-\hat{\pi}(i)]}\\
\\
2t_{pd}I_{33}\xi^{\gamma\gamma}(i)+[\varepsilon_{p}-\varepsilon_{d}-2t_{pd}g(I_{22}^{-1}+I_{33}^{-1})]I_{22}^{-1}b\xi(i)+2t_{pd}I_{33}\eta^{\gamma\gamma}(i)\\
+[\varepsilon_{d}-\varepsilon_{p}+U+2t_{pd}g(I_{22}^{-1}+I_{33}^{-1})]I_{33}^{-1}g\eta(i)+[\varepsilon_{p}-\mu-2t_{pd}g(I_{22}^{-1}+I_{33}^{-1})]\hat{\pi}(i)\\
+t_{pd}\kappa(i)+[2(\varepsilon_{d}-\varepsilon_{p})+U]\theta(i)+2t_{pp}[I_{33}p^{\gamma\lambda}(i)+\rho(i)]
\end{array}\right)\label{eq15}
\end{equation}
where the following new higher-order composite fields appear 
\[
\begin{split}\kappa(i) & =\sigma^{\mu}[d^{\dag}(i)\sigma_{\mu}p^{\gamma}(i)p^{\gamma}(i)-p^{\gamma^{\dag}}(i)\sigma_{\mu}d(i)p^{\gamma}(i)+n_{\mu}(i)d^{\gamma\gamma}(i)]+2[p^{\gamma}(i)p^{\gamma^{\dag}}(i)d(i)-d(i)d^{\gamma\gamma}(i)^{\dag}d(i)+d(i)p^{\gamma^{\dag}}(i)p^{\gamma}(i)]\\
\theta(i) & =d(i)p^{\gamma^{\dag}}(i)d(i)\\
\rho(i) & =\frac{1}{2}\sigma^{\mu}n_{\mu}(i)p^{\gamma\lambda}(i)-d(i)p^{\gamma\lambda\dag}(i)d(i).
\end{split}
\]
\end{widetext}

\subsection{Pole approximation and Green's functions}

Generically, the equations of motion of any chosen basic field can
be rewritten as 
\begin{equation}
J(i)=i\frac{\partial}{\partial t}\psi(i)=\sum\limits _{\textbf{j}}\varepsilon(\textbf{i},\textbf{j})\psi(\textbf{j},t)+\delta J(i)\label{eq17}
\end{equation}
where the energy matrix $\varepsilon(\mathbf{i},\mathbf{j})$ is determined
by the condition 
\begin{equation}
\langle\{\delta J(\textbf{i},t),\psi^{\dag}(\textbf{j},t)\}\rangle=0.\label{eq18}
\end{equation}
This way of recasting the current $J(i)$ amounts to \emph{project}
this latter on the chosen basis $\psi(i)$. Accordingly, $\delta J(i)$
contains higher-order operators \emph{orthogonal} to the basis as
well as to the physics they describe. After Eq.~(\ref{eq18}), $\varepsilon(\mathbf{i},\mathbf{j})$
can be computed by 
\begin{equation}
\varepsilon(\textbf{i},\textbf{j})=\sum\limits _{\textbf{l}}m(\textbf{i},\textbf{l})I^{-1}(\textbf{l},\textbf{j})\label{eq19}
\end{equation}
where 
\begin{equation}
\begin{split}I(\textbf{i},\textbf{j}) & =\langle\{\psi(\textbf{i},t),\psi^{\dag}(\textbf{j},t)\}\rangle\\
m(\textbf{i},\textbf{j}) & =\langle\{J(\textbf{i},t),\psi^{\dag}(\textbf{j},t)\}\rangle.
\end{split}
\label{eq20}
\end{equation}
We call $I(\textbf{i},\textbf{j})$ and $m(\textbf{i},\textbf{j})$
normalization and $m$- matrices, respectively.

Let us consider the retarded Green's function (GF) 
\begin{eqnarray}
G(i,j) & = & \langle R[\psi(i)\psi^{\dag}(j)]\rangle\label{eq21}\\
 & = & \frac{ia^{2}}{(2\pi)^{2}}\int\limits _{\Omega_{B}}d^{2}\textbf{k}d\omega e^{i\textbf{k}(\textbf{i}-\textbf{j})-i\omega(t_{i}-t_{j})}G(\textbf{k},\omega)\nonumber 
\end{eqnarray}
After Eq.~(\ref{eq17}), $G(i,j)$ satisfies the following equations
of motion 
\begin{equation}
\begin{split}i & \frac{\partial}{\partial t}G(i,j)=i\delta(t_{i}-t_{j})I(\textbf{i},\textbf{j})\\
 & +\sum\limits _{\textbf{l}}\varepsilon(\textbf{i},\textbf{l})G(\textbf{l},t_{i},\textbf{j},t_{j})+\langle R[\delta J(i)\psi^{\dag}(j)]\rangle.
\end{split}
\label{eq22}
\end{equation}
In the pole approximation, we neglect the last term, i.e. the higher-order
propagator, in Eq.~(\ref{eq22}). Then, in momentum space, the GF
satisfies the equation 
\[
[\omega-\varepsilon(\textbf{k})]G(\textbf{k},\omega)=I(\textbf{k}).
\]
The general solution of this equation reads as 
\begin{equation}
G(\textbf{k},\omega)=\sum\limits _{n=1}^{4}\frac{\sigma^{(n)}(\textbf{k})}{\omega-E_{n}(\textbf{k})+i\delta}\label{eq23}
\end{equation}
where $E_{n}(\textbf{k})$ are the eigenvalues of the energy matrix
$\varepsilon(\textbf{k})$, the spectral density matrices $\sigma^{(n)}(\textbf{k})$
can be computed as 
\begin{equation}
\sigma_{ab}^{(n)}(\textbf{k})=\sum\limits _{c=1}^{4}\Omega_{an}^{\phantom{-1}}(\textbf{k})\Omega_{nc}^{-1}(\textbf{k})I_{cb}^{\phantom{-1}}(\textbf{k}).\label{eq24}
\end{equation}
and the matrix $\Omega(\textbf{k})$ contains the eigenvectors of
$\varepsilon(\textbf{k})$ as columns.

The correlation functions (CFs) $C_{ab}(i,j)=\langle\psi_{a}(i)\psi_{b}^{\dag}(j)\rangle$
can be easily determined in terms of the GF by means of the spectral
theorem and have the general expression 
\begin{equation}
C_{ab}(\textbf{k},\omega)=\sum\limits _{n=1}^{4}[1+T_{n}(\textbf{k})]\sigma_{ab}^{(n)}(\textbf{k})\delta(\omega-E_{n}(\textbf{k})),\label{eq25}
\end{equation}
with $T_{n}(\textbf{k})=\tanh(\beta E_{n}(\textbf{k})/2)$.

Equations~(\ref{eq23}) and (\ref{eq25}) clearly show that the GF
and the CFs can be expressed in terms of the normalization matrix
$I(\textbf{k})$ and the $m$-matrix $m(\textbf{k})$ only. These
latter clearly acquire a central role in the theory and their determination
is the most relevant issue to be addressed in the following. The expressions
of $I(\textbf{k})$ and $m(\textbf{k})$ are reported in Appendices~\ref{app:basisI}
and \ref{app:basisII}, where it is shown that they depend on a set
of parameters, which are static correlation functions of composite
operators. Some of these operators belong to the chosen basis and
the related correlation functions can be easily computed self-consistently
through Eq.~(\ref{eq25}). Other operators are composite fields of
higher order, not belonging to the chosen basis, and their correlation
functions must be evaluated some other way. This crucial aspect of
the COM framework will be considered in detail in the next two subsections,
where the self-consistent schemes of calculations related to the two
possible choices of operatorial basis given above will be presented.

\subsection{Self-consistency schemes}

\subsubsection{Scheme I}

In this subsection, we report a self-consistent scheme to be used
to compute $I(\textbf{k})$ and $m(\textbf{k})$ for the first choice
of basis reported above. According to the expressions given in Appendix~\ref{app:basisI},
the normalization matrix $I(\textbf{k})$ depends on five parameters
$n_{d},b,c,\hat{a}_{s},\chi_{s}$. The first four of these parameters
can be fixed by means of the self-consistent equations 
\begin{equation}
\begin{split}n_{d} & =2[1-C_{22}-2C_{23}-C_{33}]\\
b & =\langle p^{\gamma}(i)\eta^{\dag}(i)\rangle=C_{13}^{\gamma}\\
c & =\langle p^{\gamma}(i)\xi^{\dag}(i)\rangle=C_{12}^{\gamma}\\
\hat{a}_{s} & =4\left(C_{14}^{\gamma}+3\frac{{C_{12}^{\gamma}}^{2}}{I_{22}}+3\frac{{C_{13}^{\gamma}}^{2}}{I_{33}}\right)-6\left(\frac{n_{d}}{2}-C_{33}\right).
\end{split}
\label{eq26}
\end{equation}
where 
\begin{equation}
\begin{split}C_{ab} & =\langle\psi_{a}(i)\psi_{b}^{\dag}(i)\rangle\\
C_{ab}^{\kappa} & =\langle\psi_{a}^{\kappa}(i)\psi_{b}^{\dag}(i)\rangle\\
C_{ab}^{\kappa\rho} & =\langle\psi_{a}^{\kappa\rho}(i)\psi_{b}^{\dag}(i)\rangle
\end{split}
\label{eq27}
\end{equation}
for arbitrary projectors $\kappa$ and $\rho$. In the matrix $m(\textbf{k})$,
there appear four new parameters: $\mu,\chi_{s}^{\beta},m_{44}^{(0)},m_{44}^{(\alpha)}$.
These parameters, together with $\chi_{s}$, can be determined by
means of the algebra constraints \cite{Theory,COM_in_book} dictated
by the local \emph{contractions} of the composite operators belonging
to the basis, which require 
\begin{equation}
n_{T}=2[2-C_{11}-C_{22}-2C_{23}-C_{33}]\label{eq288}
\end{equation}
\begin{equation}
\begin{split}C_{23} & =0\\
C_{24} & =3C_{12}^{\gamma}-3c\frac{C_{22}}{I_{22}}\\
C_{34} & =-3b\frac{C_{33}}{I_{33}}.
\end{split}
\label{eq28}
\end{equation}
$n_{T}=n_{d}+n_{p}$ is the total number of electrons per site. By
means of a decoupling procedure, the last parameter, $\chi_{s}^{\beta}$,
can be expressed in terms of known correlation functions 
\begin{equation}
\chi_{s}^{\beta}\approx-2\langle d^{\beta}(i)d^{\dag}(i)\rangle^{2}=-2\left(C_{22}^{\beta}+2C_{23}^{\beta}+C_{33}^{\beta}\right)^{2}.\label{eq29}
\end{equation}
By means of Eq.~(\ref{eq25}), Eqs.~(\ref{eq26}-\ref{eq29}) constitute
a set of nine coupled self-consistent equations, which will determine
the nine internal parameters. The knowledge of these parameters will
allow us to calculate various properties of the system.

As already reported in \cite{Fiorentino_01}, in some regions of the
parameters space, the Pauli conditions (\ref{eq28}) may be too restrictive
owing to the approximate nature of the solution and the system could
find it very difficult to properly adjust itself and fulfill all conditions.
Accordingly, as an alternative method, we keep (\ref{eq288}), which
mainly fixes the chemical potential, and use a decoupling procedure
to calculate $\chi_{s}$ and the higher-order correlators appearing
in $m_{44}^{(0)}$ and $m_{44}^{(\alpha)}$ (see Appendix~\ref{app:basisI}).
This procedure leads to the following set of self-consistent equations
\begin{equation}
\begin{array}{llll}
 & b_{s} & = & C_{24}^{\alpha}+C_{34}^{\alpha}+3\frac{c}{I_{22}}\left(C_{22}^{\alpha}+C_{23}^{\alpha}\right)+3\frac{b}{I_{33}}\left(C_{23}^{\alpha}+C_{33}^{\alpha}\right)\\
 & a_{s\lambda} & = & C_{14}^{\gamma\lambda}\\
 & c_{\lambda} & = & C_{12}^{\gamma\lambda}\\
 & b_{\lambda} & = & C_{13}^{\gamma\lambda}\\
 & D & = & \frac{n_{d}}{2}-C_{33}\\
 & f & = & -C_{11}^{\gamma\gamma}\left(C_{12}^{\gamma}+C_{13}^{\gamma}\right)\\
 & d_{s} & = & \left(C_{12}^{\gamma\alpha}+C_{12}^{\gamma\alpha}\right)\left(C_{22}^{\alpha}+2C_{23}^{\alpha}+C_{33}^{\alpha}\right)\\
 & \chi_{s} & = & -2\left(C_{22}^{\alpha}+2C_{23}^{\alpha}+C_{33}^{\alpha}\right)^{2}\\
 & \chi_{s}^{\beta} & = & -2\left(C_{22}^{\beta}+2C_{23}^{\beta}+C_{33}^{\beta}\right)^{2}
\end{array}\label{eq30}
\end{equation}
that supplements Eqs.~(\ref{eq26}) and closes the self-consistent
scheme necessary to compute the relevant correlation functions and
properties of the system.

\subsubsection{Scheme II}

In this subsection, we report a self-consistent scheme to be used
to compute $I(\textbf{k})$ and $m(\textbf{k})$ for the second choice
of basis reported above. According to the expressions reported in
Appendix~\ref{app:basisII}, the normalization matrix $I(\textbf{k})$
depends on five internal parameters $n_{d},a,g,f_{s},p$. The first
three of these latter can be fixed by means of the self-consistent
equations 
\begin{equation}
\begin{split}n_{d} & =\langle d^{\dag}(i)d(i)\rangle=2[1-C_{22}-2C_{23}-C_{33}]\\
g & =C_{12}^{\gamma}-C_{13}^{\gamma}\\
a & =1-C_{11}^{\gamma\gamma}\left(1-n_{d}\right)-2C_{14}^{\gamma}-2b\frac{C_{12}^{\gamma}}{I_{22}}+2b\frac{C_{13}^{\gamma}}{I_{33}}
\end{split}
\label{eq31}
\end{equation}
In the matrix $m(\textbf{k})$, there appear four new parameters:
$\mu,m_{44}^{(0)},m_{44}^{(\alpha)},m_{44}^{(\beta)}$. The parameters
$\mu,p,m_{44}^{(0)},m_{44}^{(\alpha)}$ can be determined by means
of the algebra constraints~(\ref{eq288}) and (\ref{eq28}). Using
a decoupling procedure, the two parameters $f_{s}$ and $m_{44}^{(\beta)}$
can be expressed in terms of known correlation functions as 
\begin{equation}
\begin{split}f_{s} & \approx4C_{14}^{\gamma}+4b\frac{C_{12}^{\gamma}}{I_{22}}-4b\frac{C_{13}^{\gamma}}{I_{33}}+6\left(C_{12}^{\gamma}+C_{13}^{\gamma}\right)^{2}\\
m_{44}^{(\beta)} & \approx I_{33}^{3}-\left[C_{22}^{\beta}+2C_{23}^{\beta}+C_{33}^{\beta}\right]^{2}.
\end{split}
\label{eq33}
\end{equation}

\section{Results\label{sec:results}}

\begin{figure}
\includegraphics[width=8cm]{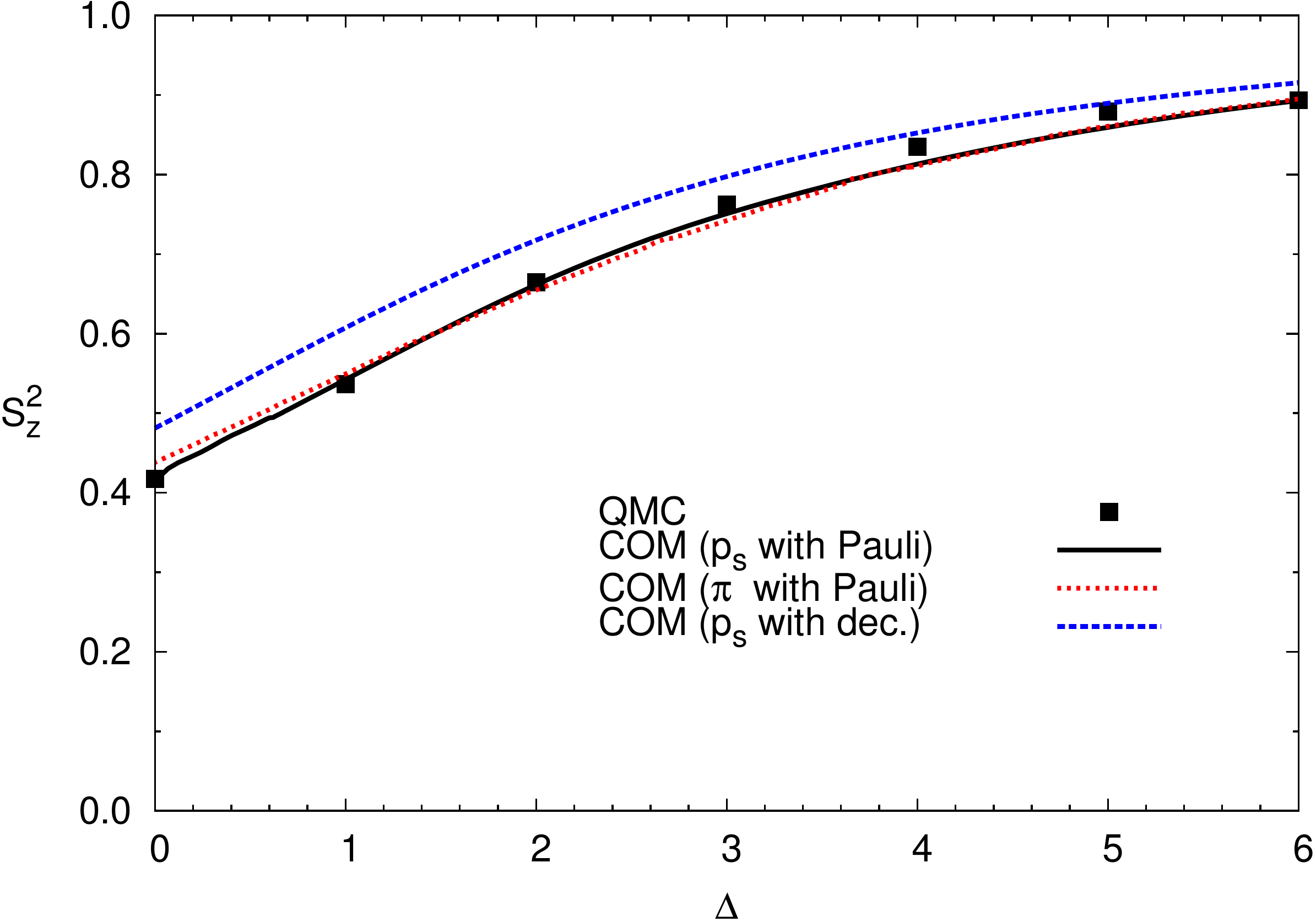} \caption{\label{fig_fio_1} (Color online) The squared local magnetic moment
per site of $d$ electrons $S_{z}^{2}$ as a function of $\Delta$
for $U=6$, $n_{T}=3$ and $T=1/3$ ($t_{pp}=0$). The solid black
line refers to the COM results obtained using the field $p_{s}$ in
the operatorial basis and the algebra constraints coming from the
Pauli principle in the self-consistent scheme, the dashed blue to
the field $p_{s}$ and the decoupling, and the dotted red line to
$\hat{\pi}$ and the Pauli principle. The black squares are qMC data
from \cite{Dopf_90}.}
\end{figure}

\begin{figure}
\includegraphics[width=8cm]{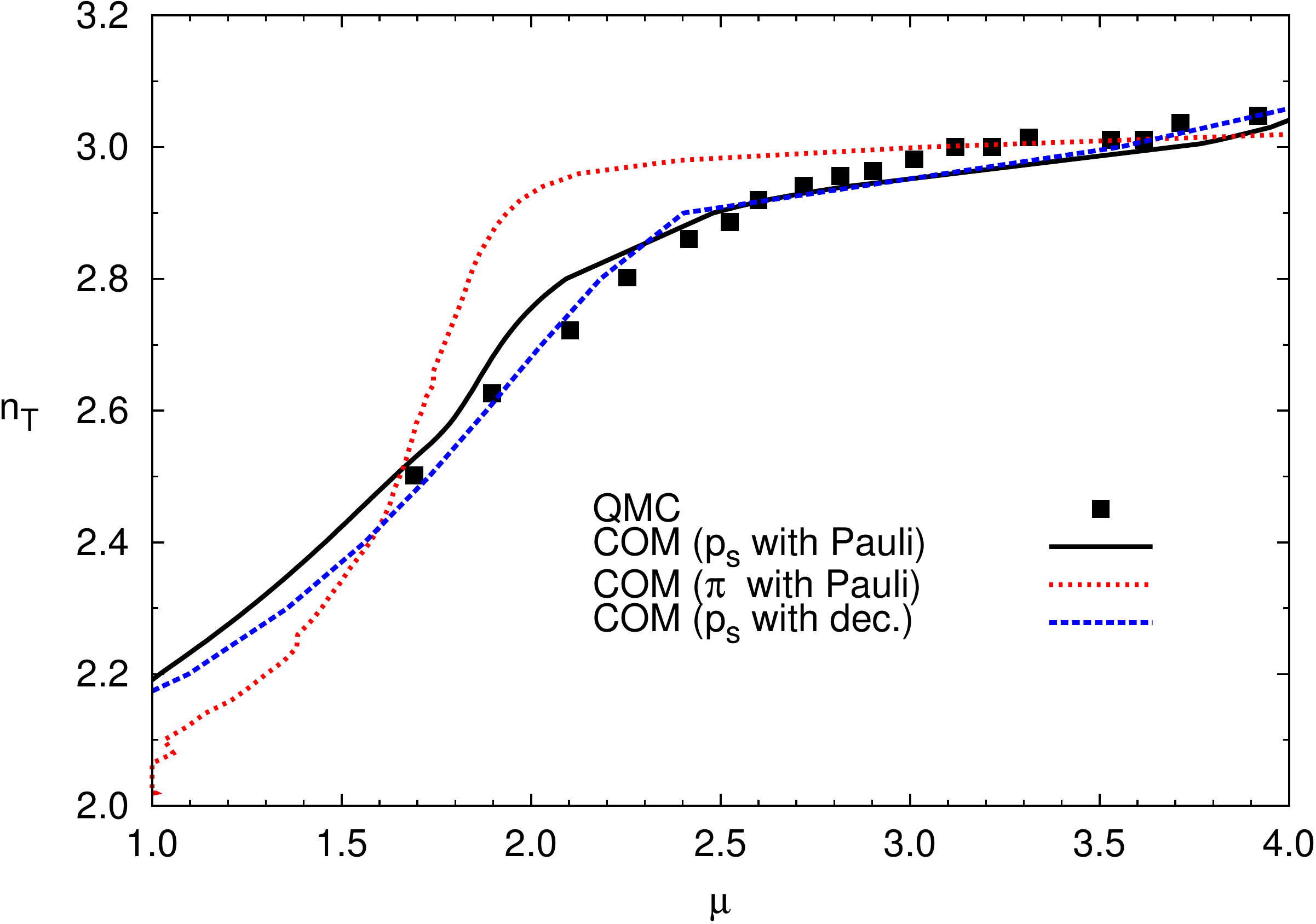} \caption{\label{fig_fio_5} (Color online) The total filling $n_{T}$ versus
the chemical potential $\mu$ is reported for $U=6$, $\Delta=4$
and $T=1/3$ ($t_{pp}=0$). The solid black line refers to the COM
results obtained using the field $p_{s}$ in the operatorial basis
and the algebra constraints coming from the Pauli principle in the
self-consistent scheme, the dashed blue to the field $p_{s}$ and
the decoupling, and the dotted red line to $\hat{\pi}$ and the Pauli
principle. The black squares are qMC data from \cite{Dopf_90}.}
\end{figure}

\subsection{Comparison with numerical simulations\label{sec:results:cfr}}

In this Section, we compare the results obtained within the COM framework
with those available in the literature, coming from a numerical analysis
performed by means of qMC method~\cite{Dopf_90}. As it has been
already shown in~\cite{Fiorentino_01}, the agreement within the
Emery model between the numerical and the COM results is excellent.
In particular, the computational schemes related to the first choice
of the basis (both the one involving the algebra constraints coming
from the Pauli principle and the one exploiting the decoupling procedure)
give an excellent agreement with qMC for several quantities ($S_{z}^{2}$,
$\mu$, band occupations, etc). In the present paper, we want to test
the second possible choice of the basis and its self-consistent scheme
in order to asses the overall stability and efficiency of the COM
framework for the Emery model. Given the very comprehensive comparative
analysis performed in~\cite{Fiorentino_01} between COM and available
numerical results, we have here limited our comparative analysis to
just two quite relevant cases. Anyway, it is worth noticing that the
second choice of the basis returns results in excellent agreement
with the numerical ones also for all other quantities analyzed in~\cite{Fiorentino_01}
using the first choice. Hereafter, all energies are given in units
of $t_{pd}$ and measured with respect to the atomic level $\varepsilon_{p}=0$.

In Fig.~\ref{fig_fio_1}, we report the squared local magnetic moment
per site of $d$ electrons $S_{z}^{2}=\frac{1}{4}\left\langle \left(n_{\uparrow}^{d}(\mathbf{i})-n_{\downarrow}^{d}(\mathbf{i})\right)^{2}\right\rangle $
as a function of $\Delta$ for $U=6$, $n_{T}=3$ and $T=1/3$. In
the paramagnetic case, $S_{z}^{2}$ can be expressed~\cite{Fiorentino_01}
through the double occupancy of $d$ electrons $D$ and the number
operator for $d$ electrons $n_{d}$ as $S_{z}^{2}=n_{d}-2D$. $D$
can be, in turn, computed directly in terms of correlation functions
involving elements of the chosen operatorial basis: $D=\frac{n^{d}}{2}-C_{33}$.
It is evident that the agreement with qMC result is very good in all
three cases, especially when one uses the algebra constraints embedding
the Pauli principle. $S_{z}^{2}$ takes the smallest value when $\Delta$
approaches zero as, in this case, the $p$ level and the $d$ upper
Hubbard subband coincide and the strong hybridization enhances the
$d$ electron double occupancy. On the other hand, when $\Delta$
becomes larger than $U$ the system moves from a charge-transfer to
a Mott-Hubbard insulator and $S_{z}^{2}$ becomes independent from
$\Delta$.

In Fig.~\ref{fig_fio_5}, the dependence of the total filling $n_{T}$
on the chemical potential $\mu$ is reported for $U=6$, $\Delta=1$
and $T=1/3$. In this case too, we find a good agreement between COM
results and qMC ones: the better agreement is reached for the solution
having the field $p_{s}$ in the operatorial basis and the decoupling
in the self-consistent scheme. The behavior of the chemical potential
close to half filling ($n_{T}=3$) -- the curve if not completely
flat is just very little tilted because of the quite high value of
the temperature -- shows clear evidences of the opening of a gap.

It is worth noticing that the COM formulation is fully self-consistent
and no adjustable parameter is used. Different choices of the operatorial
basis, and of the related self-consistent scheme, return qualitatively
similar results and definitely good and absolutely comparable benchmarks
with respect to the numerical results. In the following, we will mainly
use the solution having the field $p_{s}$ in the operatorial basis
and the decoupling in the self-consistent scheme since this is the
option that provides the higher numerical stability.

\begin{figure}
\includegraphics[width=8cm]{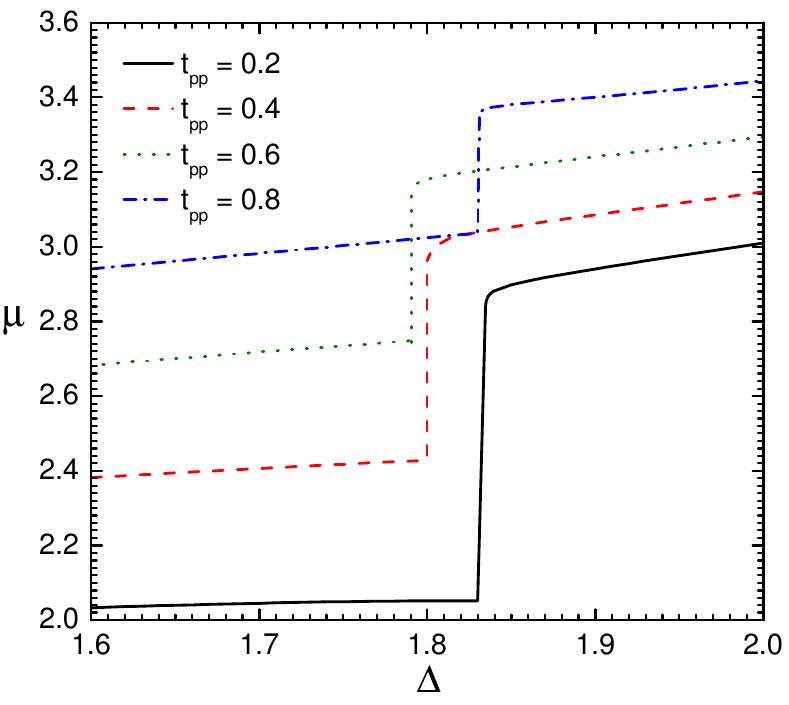} \caption{\label{fig_MIT33} (Color online) The chemical potential $\mu$ as
a function of $\Delta$ for $U=6$, $T=0.01$, $n_{T}=3$ and various
values of $t_{pp}$.}
\end{figure}

\begin{figure}
\includegraphics[width=8cm]{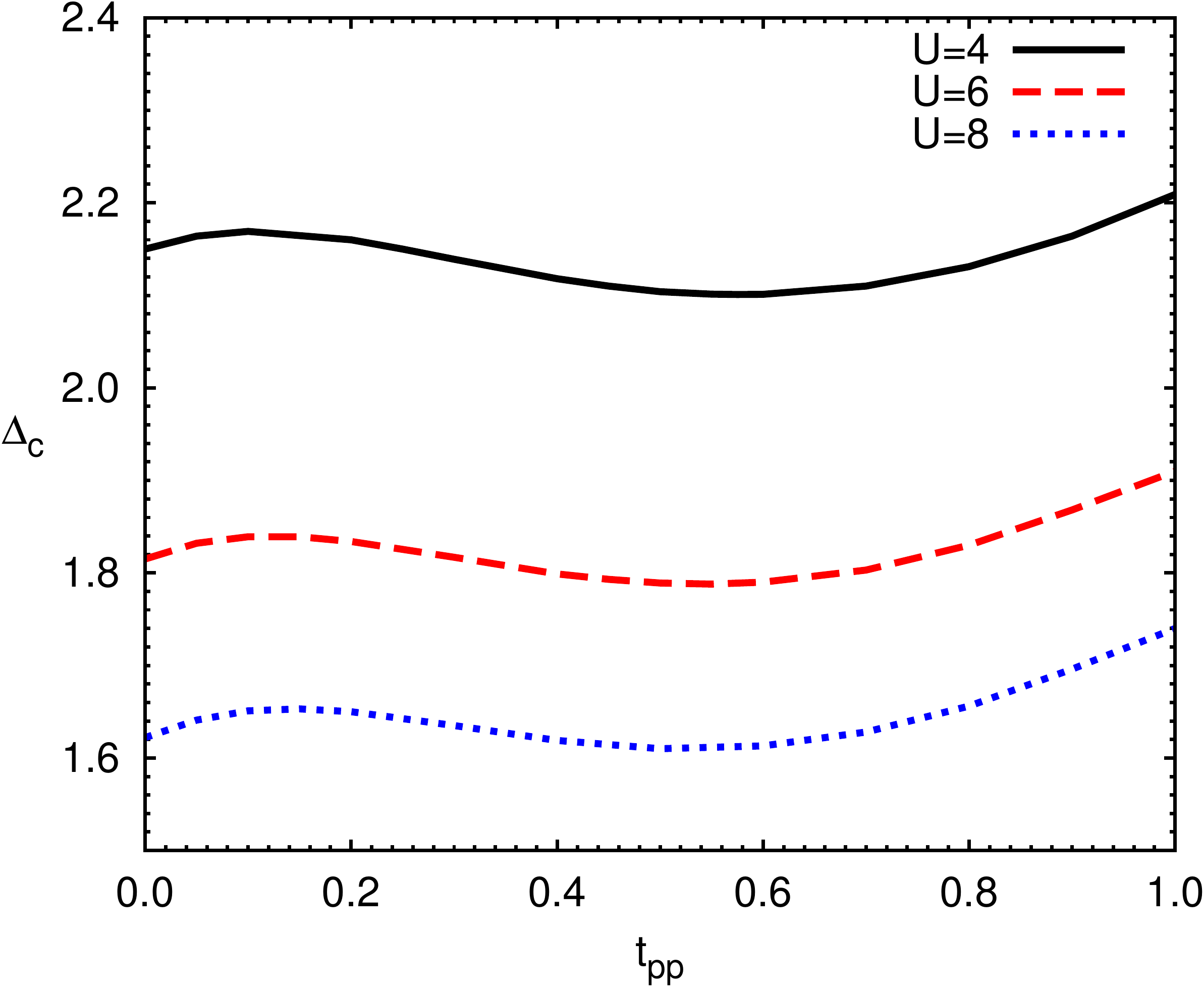} \caption{\label{fig_MIT2} (Color online) The critical value of $\Delta$,
$\Delta_{c}$, as a function of $t_{pp}$ for $T=0.01$, $n_{T}=3$
and various values of $U$.}
\end{figure}

\begin{figure*}
\includegraphics[width=8cm]{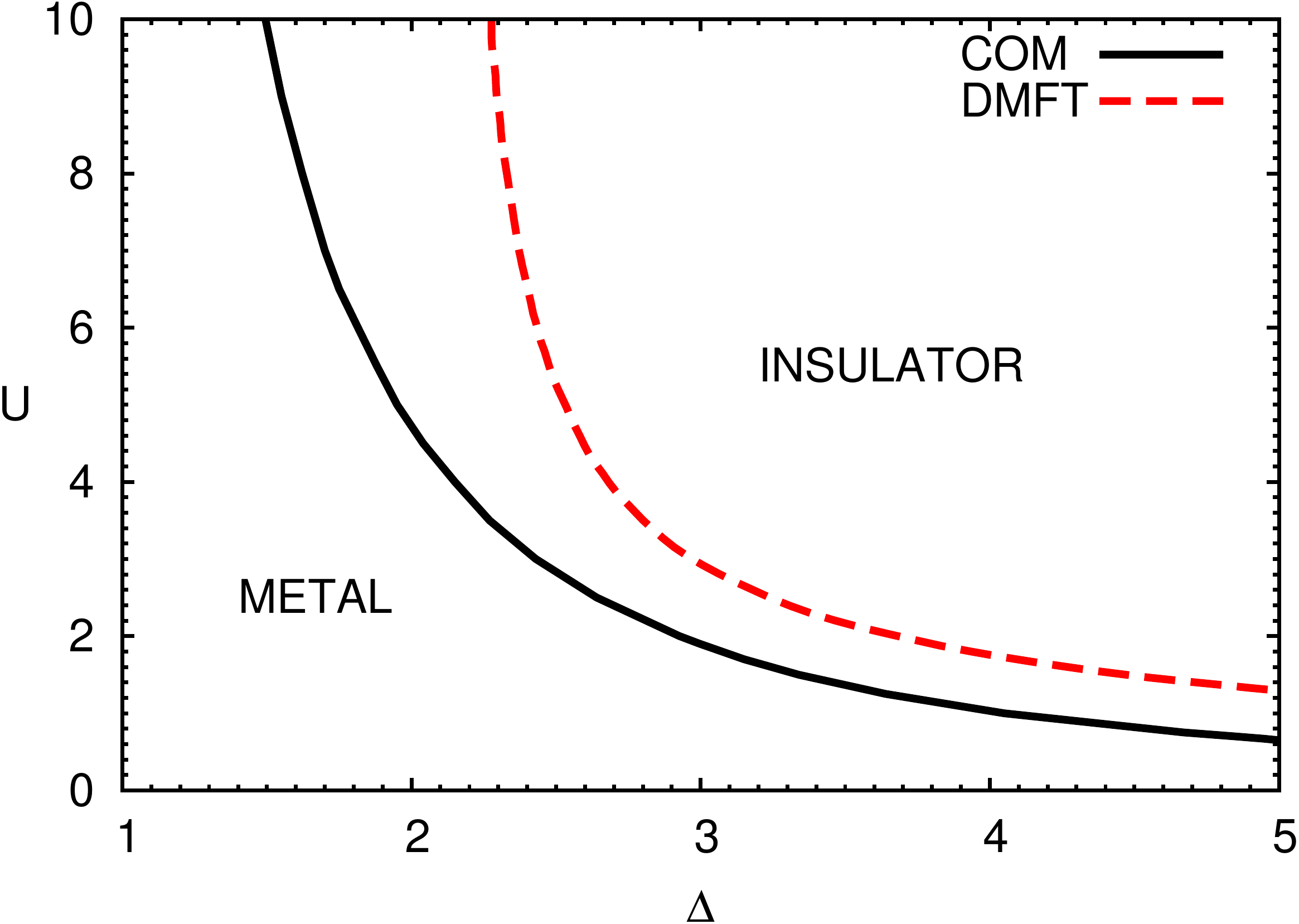} \hspace*{\fill}\includegraphics[width=8cm]{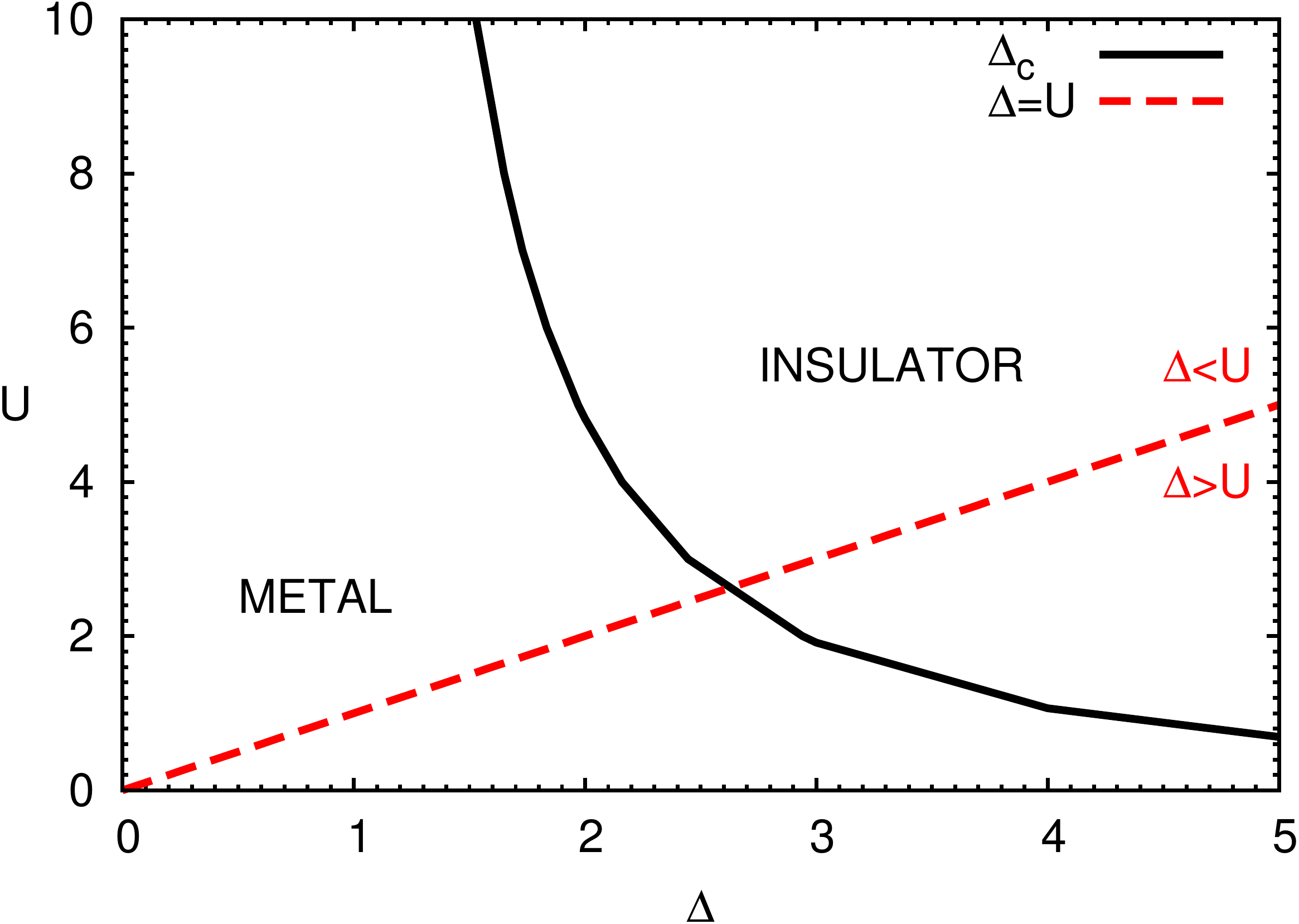}
\caption{\label{fig_MIT1} (Color online) Left panel: MIT phase diagram of
the Emery model for $T=0.01$, $n_{T}=3$ and $t_{pp}=0$: black solid
line refers to COM results, dashed red line to DMFT ones \cite{Ono_01}.
Right panel: MIT phase diagram of the Emery model for phase diagram
for $T=0.01$, $n_{T}=3$ and $t_{pp}=0.2$ within COM (solid black
line). The dashed red line is a guide to the eye and separates the
charge-transfer and Mott-Hubbard regimes in the insulator region.}
\end{figure*}

\subsection{Metal-Insulator Transition\label{sec:results:mit}}

The metal-insulator transition (MIT) has been observed in many transition-metal
oxides with more or less exotic physical properties~\cite{Mott_90,Imada_98}.
MIT has been largely studied in the single-band Hubbard model~\cite{Imada_98,Mancini_00b},
whose peculiar behavior gives the name to one of the fundamental types
of MIT (paramagnetic, homogenous, due to the strong local Coulomb
repulsion). According to \cite{Zaanen_85}, the Emery model is instead
the stage for two different types of MIT corresponding to two different
regimes mainly ruled by the relationship between $U$ and $\Delta$.
In particular, if $U>\Delta$ ($\varepsilon_{p}>\varepsilon_{d}$),
the system is said to be in the charge-transfer (CT) insulating regime
and the energy gap is roughly given by $\Delta$. On the contrary,
if $U<\Delta$ ($\varepsilon_{p}<\varepsilon_{d}$)$ $, the system
is said to be in the Mott-Hubbard (MH) regime and the energy gap is
roughly given by $U$. This is easily understandable in terms of the
relative positions of the $p$ level ($\varepsilon_{p}$) and of the
two $d$ type Hubbard sub-bands ($\varepsilon_{d}$ and $\varepsilon_{d}+U$).
Given that $\varepsilon_{p},\varepsilon_{d}<\varepsilon_{d}+U$, i.e.
given that the first unoccupied level always belongs to the upper
sub-band $\varepsilon_{d}+U$, we can have that the last occupied
level either belongs to $\varepsilon_{p}$ in a CT or to $\varepsilon_{d}$
in a MH. The size of the gap follows immediately.

In the present paper, the study of the MIT in the Emery model has
a two-fold aim: on one hand, we are interested in the mere occurrence
of the MIT in the model and in its characterization through the analysis
of its properties; on the other hand, our findings will shed some
more light onto the puzzling physics of cuprates. In particular, we
will discuss the relationship between the ab-initio determination
of the model parameters proper for cuprates and the values of the
model parameters required by many-body treatments in order to reproduce
the experimental results.

We start from the first objective and explore the MIT in the whole
range of values of the on-site potential $U$, the charge-transfer
gap $\Delta$, and hopping integral between the oxygen orbitals $t_{pp}$.
The MIT manifests itself in various properties of the system. Some
of these properties are particularly suitable for the precise determination
of the MIT onset within the COM framework. These fully equivalent
MIT criteria are (at $T\approx0$ and $n_{T}=3$): i) the presence
of a jump in the chemical potential as a function of $\Delta$ or
$U$ and ii) the opening of a gap in the density of states (DOS) at
the Fermi level on varying $\Delta$ or $U$. The results presented
in this subsection have been obtained by using the first criterion,
namely the jump in $\mu(\Delta)$. It follows from ab-initio calculations
for single-layer cuprates~\cite{Kent_08} that typical values of
the ratio $t_{pp}/t_{pd}$ range between $0.16$ and $0.7$. As shown
in Fig.~\ref{fig_MIT33}, by varying $t_{pp}$ within this range,
we obtain a series of jumps in the dependence $\mu(\Delta)$. At the
MIT, the chemical potential jumps between the upper Hubbard sub-band
of copper main character and the band of mixed character that will
be later related to the ZRS. We can immediately see that the critical
value of $\Delta$, $\Delta_{c}$, has no monotonous behavior as a
function of $t_{pp}$. The effective dependence of $\Delta_{c}$ on
$t_{pp}$ is shown in Fig.~\ref{fig_MIT2}. It is now evident that
$t_{pp}$ does not significantly influence the overall value of $\Delta_{c}$,
once $U$ is fixed. As $t_{pp}$ increases, $\Delta_{c}$ only slowly
varies around a mean value up to $t_{pp}\approx0.8$, while for greater
values of $t_{pp}$ it starts growing quite rapidly. Interestingly,
as $U$ increases, the whole curve $\Delta_{c}(t_{pp})$ rigidly moves
downwards, almost without changing its overall shape. With respect
to the MIT, it is evidently $U/\Delta$ the main player among the
model parameters, but it is also clear that $t_{pp}$ can play a significative
role in order to fine tune the results between different materials
within the same class.

In the left panel of Fig.~\ref{fig_MIT1}, we plot the phase diagram
of the Emery model at $T=0.01$, $n_{T}=3$ and $t_{pp}=0$. The black
solid and red dashed lines mark the phase boundary separating the
metallic and the insulating regions given by COM and DMFT \cite{Ono_01},
respectively. The two critical curves are in quite good qualitative
agreement with each other. The DMFT curve lays above the COM one,
underestimating, with respect to the latter, the insulating phase,
i.e. requiring rather higher values of $\Delta$ and $U$ to realize
the MIT (see more below). In the right panel of Fig.~\ref{fig_MIT1},
we plot $\Delta_{c}$ as a function of $U$ for $T=0.01$, $n_{T}=3$
and $t_{pp}=0.2$ (black solid line). The Mott-Hubbard regime is separated
from the charge-transfer one by the red dashed line. One immediately
notice that $\Delta_{c}$ increases quite rapidly on decreasing $U$.
In the Mott-Hubbard regime, the metallic phase occurs only for small
values of $U$ ($U<2.6$). On the other hand, in the charge-transfer
regime, no metallic phase is observed for $\Delta\gtrsim2.6$. For
small values of $\Delta$, the system is metallic even in the limit
$U\to\infty$, as already pointed out in \cite{Zaanen_85,Ono_01}.
For $U>4$, the transition is always of charge-transfer type with
$\Delta\lesssim2.6$. On the other hand, it is known that several
many-body treatments fail to reproduce the insulating behavior at
half filling~\cite{Kent_08} unless $\Delta$ exceeds $3\: eV$.
Assuming a reasonable value of $t_{pd}\approx1e\: V$~\cite{Kent_08},
in our analysis we naturally obtain the charge-transfer insulator
for a wider range of $\Delta$ and $U$ without any adjustment, thus
being more consistent with the ab-initio predictions for $\Delta_{c}$
in cuprates that hardly exceed $3\: eV$~\cite{Kent_08} than other
many-body treatments that seems to underestimate the insulating phase.

\subsection{Single-particle properties and ZRS\label{sec:results:spzr}}

In this section, we analyze the single-particle properties of the
Emery model (the energy bands, their spectral weights, orbital characters,
and effective tight-binding parameters, and the density of states)
as well as the occupations per orbital and band. The purpose of this
study is (i) to fully characterize the low-energy excitations in momentum,
band and orbital, (ii) to qualitatively and quantitatively analyze
the hybridization between copper and oxygen orbitals and (iii) to
assess the validity of the ZR scenario and establish its limitations.

\begin{figure}
\includegraphics[width=8cm]{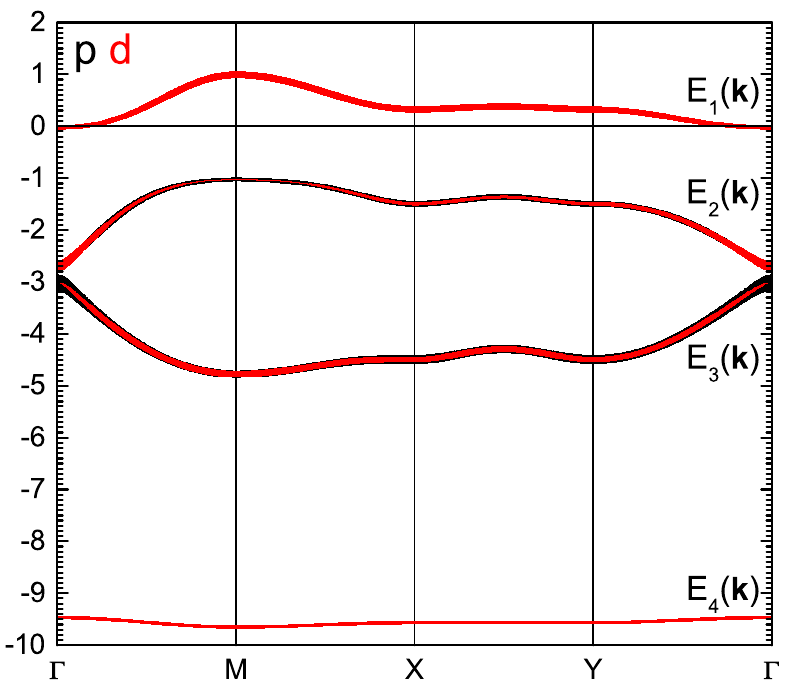} \caption{\label{fig_a1} (Color online) Electronic dispersion in momentum space
along the path $\mathrm{\Gamma=(0,0)}\to\mathrm{M}=(\pi,\pi)\to\mathrm{X=(\pi,0)}\to\mathrm{Y}=(0,\pi)\to\mathrm{\Gamma}$
for $n_{T}=3$, $\Delta=2$, $t_{pp}=0.2$, $T=0.01$, and $U=6$.
The width of the bands is proportional to their spectral weights as
functions of the momentum per orbital character: the $p$- and $d$-
orbital characters are depicted with black and red colors, respectively.}
\end{figure}

\begin{figure}
\includegraphics[width=8cm]{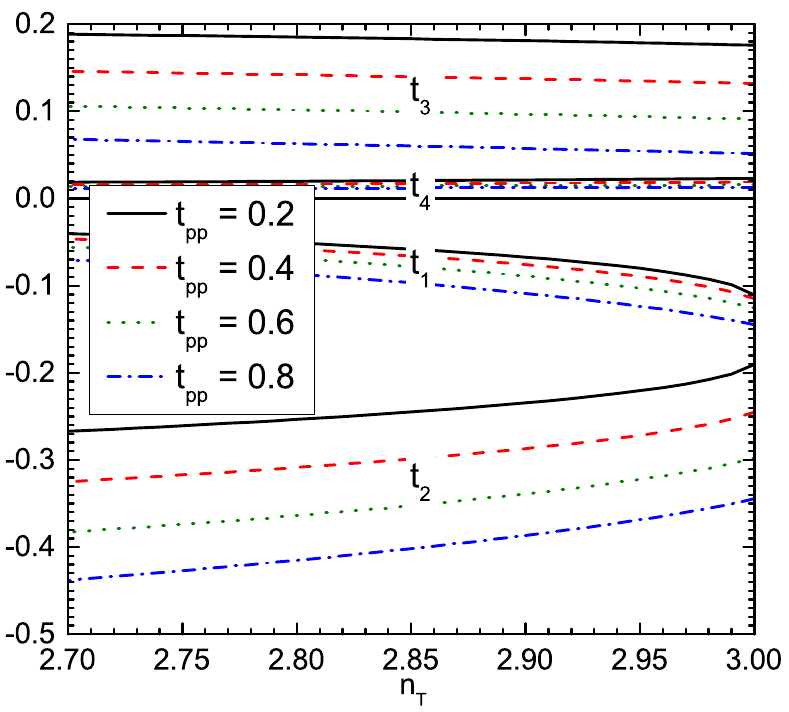}

\includegraphics[width=8cm]{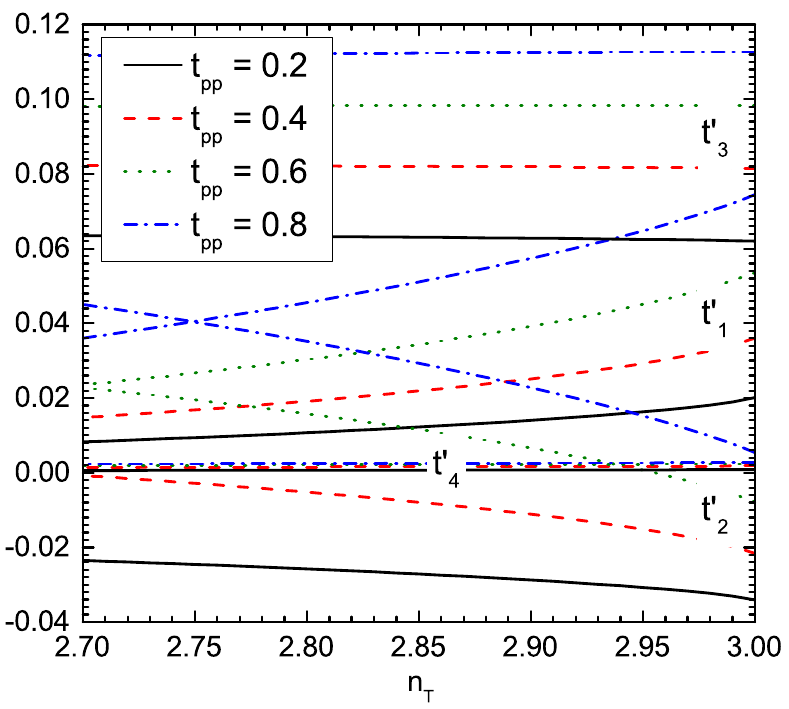} \caption{\label{fig_a7} (Color online) Total filling ($n_{T}$) dependence
of the effective nearest-neighbor ($t_{n}$, top panel) and next-nearest-neighbor
($t_{n}^{\prime}$, bottom panel) hopping integrals of the four bands
$E_{n}(\mathbf{k})$ at $\Delta=\Delta_{c}$, $U=6$, $T=0.01$ and
various values of $t_{pp}$.}
\end{figure}

\subsubsection{Bands}

The adopted four-pole approximation obviously returns a four-band
structure ($E_{n}(\mathbf{k})$ with $n=1,\ldots,4$) for the electronic
dispersion of the Emery model, as it can be clearly seen in Fig.~\ref{fig_a1},
where the latter is reported as a function of the momentum along the
path $\mathrm{\Gamma=(0,0)}\to\mathrm{M}=(\pi,\pi)\to\mathrm{X=(\pi,0)}\to\mathrm{Y}=(0,\pi)\to\mathrm{\Gamma}$
for $n_{T}=3$, $\Delta=2$, $t_{pp}=0.2$, $T=0.01$, and $U=6$.
The fictitious width of the four bands is proportional to their spectral
weights as functions of the momentum per orbital character: $\sigma_{p}^{(n)}(\textbf{k})=\sigma_{11}^{(n)}(\textbf{k})$
and $\sigma_{d}^{(n)}(\textbf{k})=\sum_{a,b=2}^{3}\sigma_{ab}^{(n)}(\textbf{k})$.
The $p$- and $d$- orbital characters are depicted with black and
red colors, respectively. Given the center-of-mass positions of the
four bands, their dependences on momentum and their dominant orbital
characters: $E_{1}$ and $E_{4}$ can be safely assigned to the two
Hubbard sub-bands of $d$ electrons (upper and lower, respectively),
$E_{3}$ to the \emph{bare} $p$ electron level and $E_{2}$ to the
band arising from the ZR conjecture, i.e. to the dispersion related
to the one-particle removal process out of the ZRS. Accordingly, four
is exactly the minimal number of basic fields necessary to describe
these very elementary excitations, whose presence in the system is
quite well established~\cite{Zaanen_85}. Let us discuss now, one
by one, the relevant and peculiar features of these four bands.

As regards the momentum dependence, a very simple way to analyze it
just requires to Fourier transform back to real space each of the
four bands separately. It is definitely worth pointing out that this
very systematic analysis also permits to investigate in detail the
possibility to reduce the system to an effective two-band one. This
simple, but efficient, procedure returns the effective tight-binding
parameters per band as modified, with respect to the bare ones present
in the Hamiltonian, by the interactions, the hybridizations and the
related high-order real and virtual processes. As shown in Fig.~\ref{fig_a7},
$E_{1}$ and $E_{2}$ have sizable negative nearest-neighbor effective
hopping integrals ($t_{1}$ and $t_{2}$, respectively), in agreement
with the presence of a minimum at $\Gamma$ and of a maximum at $M$
(see Fig.~\ref{fig_a1}). The situation is completely reversed for
$E_{3}$ and $E_{4}$. All bands, except for $E_{4}$ that is almost
flat, have non-negligible next-nearest-neighbor (along main diagonals)
effective hopping integrals ($t_{i}^{\prime}$), in agreement with
the presence of a \emph{warp} along the $X\to Y$ direction (see Fig.~\ref{fig_a1}).
The apparent dominance (see Fig.~\ref{fig_a1}) of the $\alpha(\mathbf{k})$
component in the dispersion (among the 2D cubic harmonics) is reflected
in the prevalence of $t_{n}$ on longer-distance hopping integrals.
Given the hopping and the hybridization terms present in the Hamiltonian
(\ref{eq:ham}), the $\alpha(\mathbf{k})$ component of the dispersion
is dynamically generated by the second-order process describing the
indirect hopping between nearest-neighbor copper sites involving an
intermediate oxygen site (i.e. $t_{n}\propto t_{pd}^{2}$ and $\gamma^{2}(\textbf{k})=1-\alpha(\mathbf{k})$).
Similarly, it is the Hamiltonian term responsible for the direct hopping
between nearest-neighbor oxygens to induce a finite value of the next-nearest-neighbor
(along main diagonals) effective hopping integrals ($t_{i}^{\prime}$).
It also induces the related $\beta(\mathbf{k})$ component of the
dispersion, which drives the \emph{warp} along the $X\to Y$ direction.
These two occurrences are based on the third-order process describing
the indirect hopping between next-nearest-neighbor (along main diagonals)
copper sites involving two intermediate nearest-neighbor oxygen sites
(i.e. $t'_{n}\propto t_{pd}t_{pp}t_{pd}$ and $\lambda(\textbf{k})\gamma^{2}(\textbf{k})=1-2\alpha(\textbf{k})+\beta(\textbf{k})$).
Accordingly, on top of the interactions, which play the main role
in this, also $t_{pd}$ and $t_{pp}$ enter into the redefinition
of the center-of-mass positions of the four bands. Moreover, the value
of $t_{pp}$ affects the bandwidth of the four bands as it contributes
to determine the value of $t_{n}$. Given these occurrences, it is
now clear why $t_{pp}$ is usually considered the major discriminant
among the different cuprate families or even specific materials. $E_{2}(\mathbf{k})$
features a finite next-nearest-neighbor effective hopping integral
along main axes too (not shown).

The values of all effective hopping integrals are almost completely
independent of $\Delta$ and $U$ (not shown); this occurrence opens
up the possibility to use the two bands closer to the Fermi level,
and in particular their effective hopping integrals (their \emph{tight-binding
reduction}), as constitutive elements of an effective two-band model
as suggested by the ZR conjecture. On the other hand, the dependence
of the same effective hopping integrals on the total filling $n_{T}$
is not negligible (see Fig.~\ref{fig_a7}). This latter occurrence
does not contradict the ZR conjecture as the hopping integrals of
the effective two-band model can assume effective values on varying
any external parameter (filling, temperature or any applied field).
The center-of-mass positions of the four bands, as functions of $U$
for fixed $\Delta$, do not change (not shown) except for the lower
Hubbard one ($E_{4}$). Such a behavior was foreseeable as $\Delta$
was kept constant, but only as regards the relative positions: the
constancy of the chemical potential pins the absolute positions of
the levels too. It is worth noting that the electronic dispersion
close to the Fermi level has an overall shape in very good agreement
with what found by the ab-initio calculations. 

\begin{figure}
\includegraphics[width=8cm]{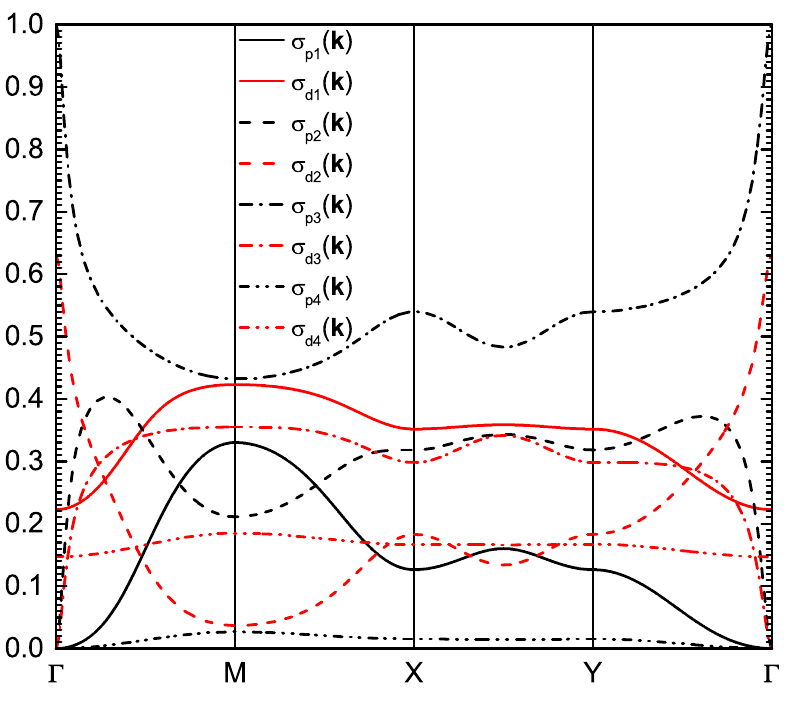}\caption{\label{fig_a3} (Color online) Spectral weights per orbital character
as functions of the momentum along the path $\mathrm{\Gamma}\to\mathrm{M}\to\mathrm{X}\to\mathrm{Y}\to\mathrm{\Gamma}$
for $n_{T}=3$, $\Delta=2$, $t_{pp}=0.2$, $T=0.01$, $U=6$. The
$p$- and $d$- orbital characters are depicted with black and red
colors, respectively.}
\end{figure}

Obviously, each of the four bands has mixed $p$- and $d$- character
as it can be clearly seen in Fig.~\ref{fig_a3}. The relative weight
of the two species ($p$- and $d$-) is quite strongly momentum dependent
and varies very much from band to band at fixed momentum. $E_{4}$
is mostly $d$-like, while $E_{1}$, $E_{2}$ and $E_{3}$ have fully
mixed character with a slight predominance of the $d$ component in
$E_{1}$ and of the $p$ component in $E_{2}$ and $E_{3}$. $E_{1}$
shows a net prevalence of the $d$ component at the $\Gamma\;(0,0)$
point, which hosts the minimum of the band, in perfect agreement with
the well established fact that the first electronic addition should
have definite $d$ character. Similarly, the ratio between the $p$-
and $d$- character ($3:2$), at the maximum of the $E_{2}$ band
(the $M(\pi,\pi)$ point), agrees very well with the evidences for
a mainly $p$ character for the first electronic removal, as also
required by the ZRS conjecture. This latter also requires an almost
identical slope in momentum space of the two components at the $M$
point, as it is apparent in Fig.~\ref{fig_a3}. In $E_{2}$, the
two components exchange their relative relevance going from the $\Gamma$
point ($d$-predominance) towards the $M$ point ($p$-predominance).
In $E_{3}$, the two components have the tendency to occupy similarly
and quite uniformly the momentum space except at the $\Gamma$ point,
where $d$-electrons are simply absent. $E_{4}$, according to its
extreme flatness, has a uniform occupation in momentum space. In conclusion,
the ZR scenario is fully supported by our findings as they show that
doping the system out of half filling ($n_{T}=3$) generates holes
in the $E_{2}$ band at $(\pi,\pi)$ with a dominant $p$-character
and particles in the $E_{1}$ band at $(0,0)$ with definite $d$-character.

\subsubsection{Density of states}

\begin{figure}
\includegraphics[width=8cm]{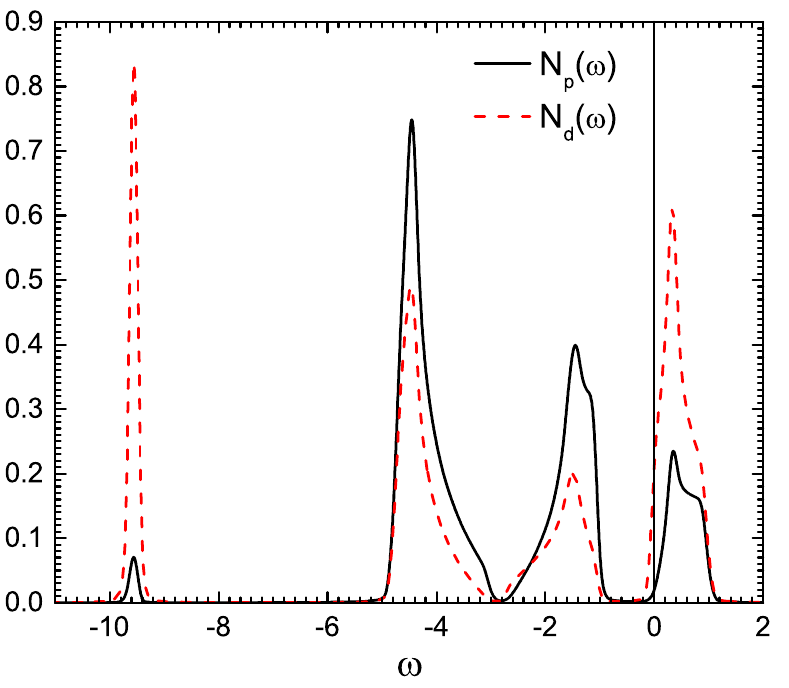}\caption{\label{fig_a4} (Color online) Densities of states per orbital character
for $n_{T}=3$, $\Delta=2$, $t_{pp}=0.2$, $T=0.01$ and $U=6$.
The $p$- and $d$- orbital characters are depicted with (solid) black
and (dashed) red colors, respectively.}
\end{figure}

The densities of states for $p$- and $d$- orbital characters can
be calculated according to the following formulas 
\begin{equation}
\begin{split}N_{p}(\omega) & =\frac{a^{2}}{(2\pi)^{2}}\int_{\Omega_{B}}d^{2}\textbf{k}\sum_{n=1}^{4}\sigma_{11}^{(n)}(\textbf{k})\delta(\omega-E_{n}(\textbf{k}))\\
N_{d}(\omega) & =\frac{a^{2}}{(2\pi)^{2}}\int_{\Omega_{B}}d^{2}\textbf{k}\sum_{n=1}^{4}\left[\sigma_{22}^{(n)}(\textbf{k})+2\sigma_{23}^{(n)}(\textbf{k})\right.\\
 & +\left.\sigma_{33}^{(n)}(\textbf{k})\right]\delta(\omega-E_{n}(\textbf{k})).
\end{split}
\end{equation}
These densities of states, presented in Fig.~\ref{fig_a4}, clearly
show the positions of the van Hove singularities together with the
marked enhancements (in particular, in $E_{1}$ and $E_{2}$) coming
from the \emph{warp} in the dispersion along the $X\to Y$ direction
due to the finite value of $t_{pp}$. The already discussed mixed
$p$- and $d$- character of the four bands is evident once more and
the strong energy dependence of the relative weight of the two species
($p$- and $d$-) is clearly visible. The very high degree of hybridization
between the two species, in particular in the energy and momentum
regions close to the Fermi surface, is another fundamental brick in
the construction of an effective theory with a reduced number of degrees
of freedom.

\subsubsection{Band and orbital occupations}

It is very interesting to analyze the distribution of the electrons
among the four bands as well as between the $p$- and $d$- orbitals.
The corresponding occupation numbers per band ($i=1,\ldots,4$) $n_{i}=n_{ip}+n_{id}$
and per orbital ($p$ or $d$) 
\[
n_{p}=\sum_{i=1}^{4}n_{ip}\qquad\qquad n_{d}=\sum_{i=1}^{4}n_{id}
\]
can be obtained properly summing up the following basic quantities
\[
\begin{split}n_{ip} & =\frac{a^{2}}{(2\pi)^{2}}\int_{\Omega_{B}}d^{2}\textbf{k}\left(1-T_{i}(\textbf{k})\right)\sigma_{11}^{(i)}(\textbf{k})\\
n_{id} & =\frac{a^{2}}{(2\pi)^{2}}\int_{\Omega_{B}}d^{2}\textbf{k}\left(1-T_{i}(\textbf{k})\right)\times\\
 & \times\left(\sigma_{22}^{(i)}(\textbf{k})+\sigma_{33}^{(i)}(\textbf{k})+2\sigma_{23}^{(i)}(\textbf{k})\right)
\end{split}
\]

\begin{figure}
\includegraphics[width=8cm]{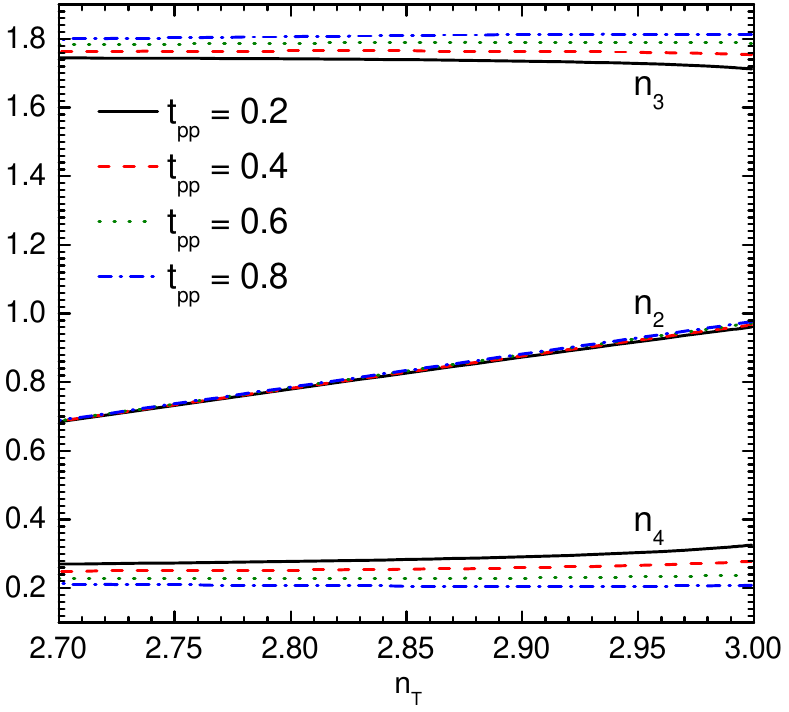} \caption{\label{fig_a6} (Color online) Occupation numbers per band $n_{2}$,
$n_{3}$ and $n_{4}$ ($n_{1}=0$) as functions of the total density
$n_{T}$ for different values of $t_{pp}$ at $\Delta=\Delta_{c}(t_{pp})$
for $T=0.01$ and $U=6$.}
\end{figure}

\begin{figure}
\includegraphics[width=8cm]{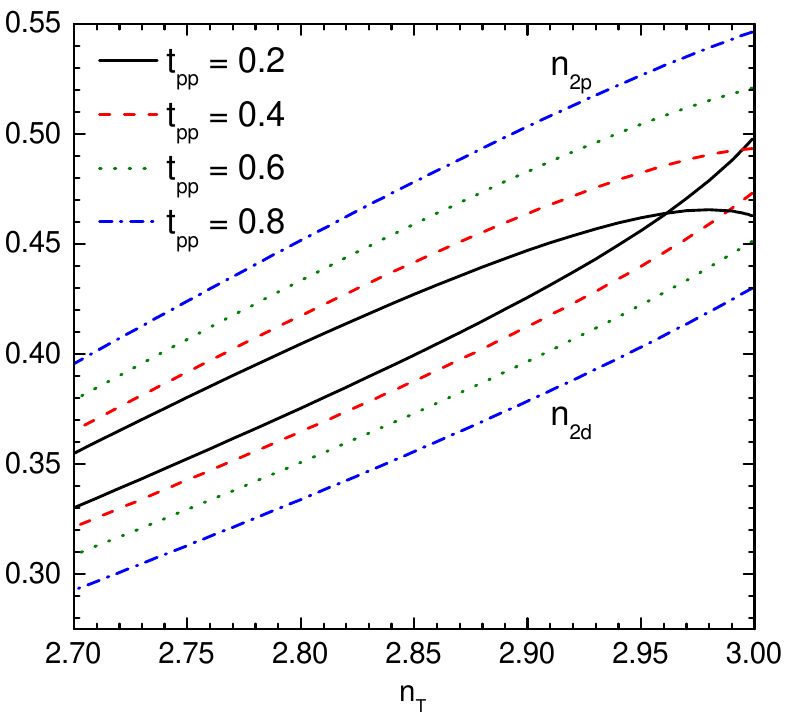} \caption{\label{fig_a8} (Color online) Occupation numbers of the $E_{2}$
band per orbital character $n_{2p}$ and $n_{2d}$ as functions of
the total density $n_{T}$ for different values of $t_{pp}$ at $\Delta=\Delta_{c}(t_{pp})$
for $T=0.01$ and $U=6$.}
\end{figure}

It is worth noting that the band structure of the Emery model determines
and is, in turn, determined by the occupations per band of the $p$-
and $d$- electrons. The analysis of such occupations per band reveals
that they are independent of $U$ (not shown), favoring the reduction
to an effective single-band model, although this is due to a compensation
of the $p$- and $d$- components within each band (not shown). In
particular, the $p$- and $d$- fillings in the band $E_{2}$ change
quite much upon varying $U$, but conserving their sum, and so the
overall filling of the band, practically constant. The reason behind
this redistribution is quite easy to understand: the on-site Coulomb
repulsion $U$ is the main control of the overall amount of double
occupancy of the $d$ component and, consequently, is also responsible
for the fine tuning of the ratio between the $p$- and $d$- components
within the ZRS.

\begin{figure}
\includegraphics[width=8cm]{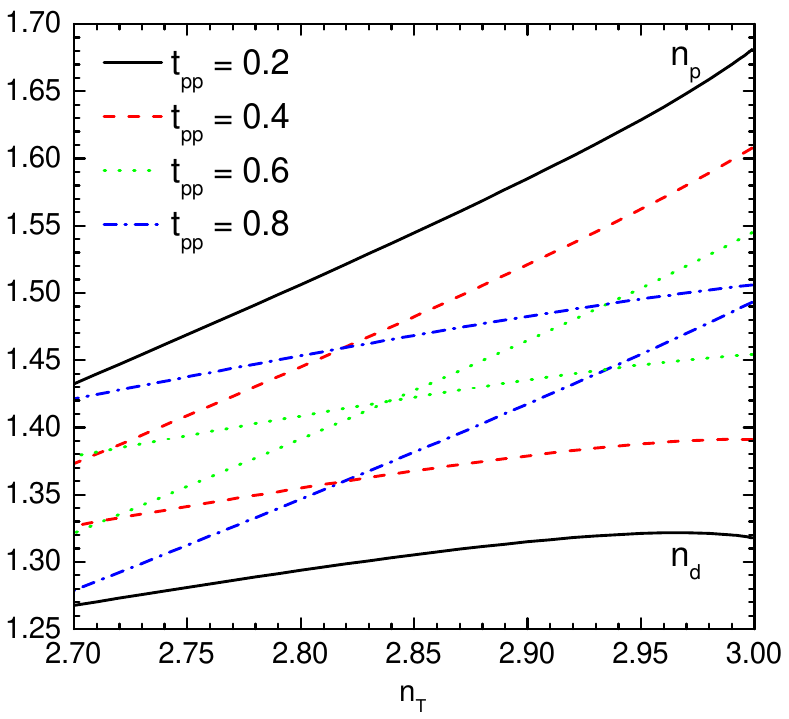} \caption{\label{fig_a9} (Color online) Occupation numbers per orbital character
$n_{p}$ and $n_{d}$ as functions of the total density $n_{T}$ for
different values of $t_{pp}$ at $\Delta=\Delta_{c}(t_{pp})$ for
$T=0.01$ and $U=6$.}
\end{figure}

\begin{figure}
\includegraphics[width=8cm]{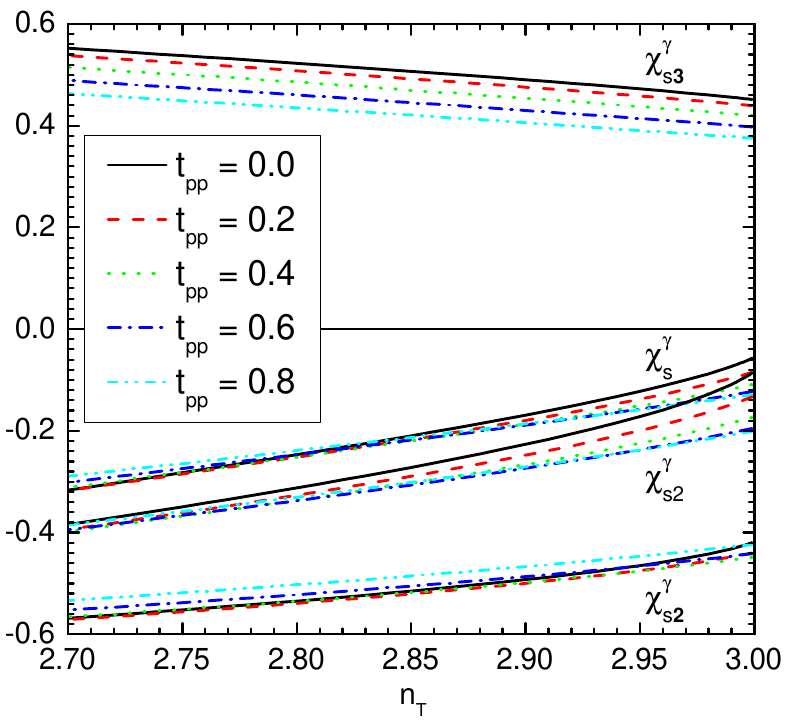} \caption{\label{fig_a} (Color online) Plaquette $p$-$d$ spin correlation
function and its per-band decomposition as functions of the total
density $n_{T}$ for different values of $t_{pp}$ at $\Delta=\Delta_{c}(t_{pp})$
for $T=0.01$ and $U=6$.}
\end{figure}

\begin{figure}
\includegraphics[width=8cm]{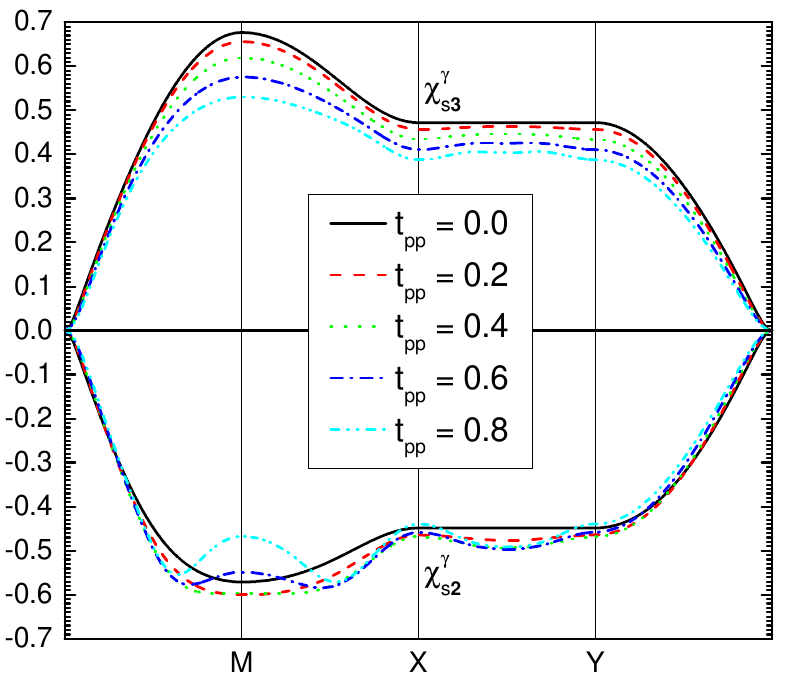} \caption{\label{fig_b} (Color online) Plaquette $p$-$d$ spin correlation
function decomposed per band as function of the momentum along the
path $\mathrm{\Gamma}\to\mathrm{M}\to\mathrm{X}\to\mathrm{Y}\to\mathrm{\Gamma}$
for $n_{T}=3^{-}$, different values of $t_{pp}$, $\Delta=\Delta_{c}(t_{pp})$,
$T=0.01$ and $U=6$.}
\end{figure}

\begin{figure}
\includegraphics[width=8cm]{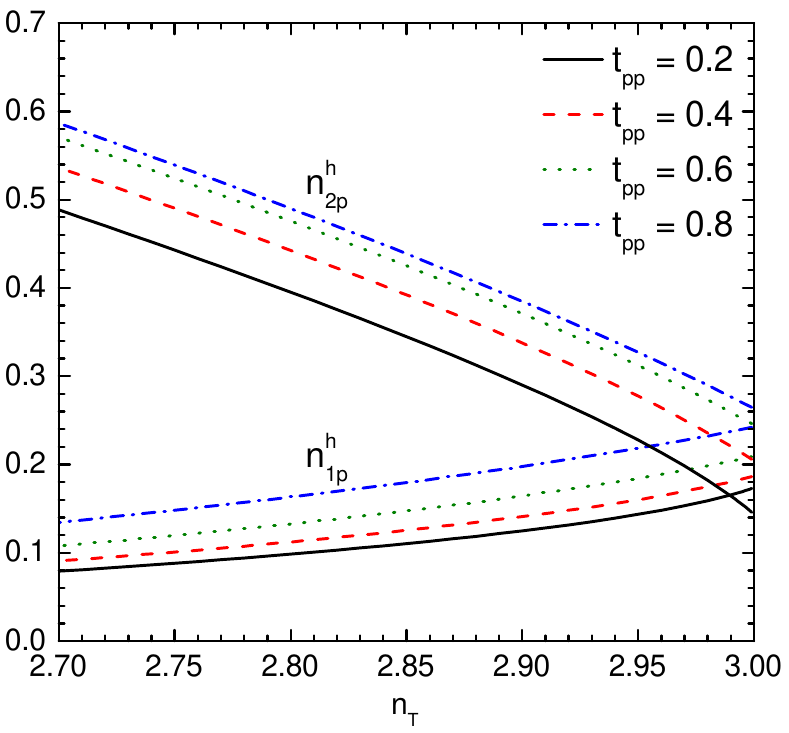} \caption{\label{fig_c} (Color online) Hole occupation numbers of the $E_{1}$
and the $E_{2}$ band per $p$ orbital character $n_{2p}$ and $n_{2d}$
as functions of the total density $n_{T}$ for different values of
$t_{pp}$ at $\Delta=\Delta_{c}(t_{pp})$ for $T=0.01$ and $U=6$.}
\end{figure}

In Fig.~\ref{fig_a6}, it is clearly shown that upon doping, the
holes go primarily into the $E_{2}$ band, while the occupation of
the other bands remains almost unchanged: the practical constancy
of the occupations of the $E_{3}$ and $E_{4}$ bands conveys the
doping only to the $E_{2}$ band, once more justifying a reduction
to an effective single-band Hubbard model. It is really remarkable
that at the MIT, marked by the vanishing of the occupation $n_{1}$
of the upper Hubbard sub-band $E_{1}$, the occupation $n_{2}$ of
the band $E_{2}$ crosses one (see again Fig.~\ref{fig_a6}). This
occurrence opens up the possibility to mime such a MIT of charge-transfer
type through a MIT of Mott-Hubbard type according to the ZR construction.
This latter construction also requires that, upon doping, the copper
and oxygen occupations of the band $E_{2}$ decrease with the same
slope: as reported in Fig.~\ref{fig_a8}, except for the doping region
really very close to the MIT or for too high values of $t_{pp}$,
this is the case. Across all four bands, the great majority of the
doping goes in the oxygen channel (see Fig.~\ref{fig_a9}) as expected~\cite{Zhang_88}.

In order to characterize further the two bands closer to the chemical
potential and with mainly $p$ character ($E_{2}$ and $E_{3}$),
we have analyzed the spin-spin correlation function between the $p$-bonding
hole projected on the plaquette and the central $d$ hole: $\chi_{s}^{\gamma}=\frac{1}{3}\left\langle \hat{p}^{\dagger\gamma}\sigma_{k}\hat{p}^{\gamma}\hat{n}_{k}^{d}\right\rangle =\frac{2}{3}\left(C_{14}^{\gamma}+\frac{3c}{I_{22}}C_{12}^{\gamma}+\frac{3b}{I_{33}}C_{13}^{\gamma}\right)$.
The hatted operators stand for their hole counterparts given by a
particle-hole transformation ($\psi_{\sigma}(i)\to(-1)^{|i|}\hat{\psi}_{\sigma}^{\dagger}(i)$).
In Fig.~\ref{fig_a}, we report $\chi_{s}^{\gamma}$ as a function
of the total density $n_{T}$ and its components per band $\chi_{sn}^{\gamma}$
in the hole representation. We also report its components per band
$\chi_{s\mathbf{n}}^{\gamma}$ computed without taking into account
the actual occupation of the related bands (i.e. without weighting
the inner momentum components by the appropriate Fermi function).
In this way, we can analyze the features of the contributions not
activated yet by the actual hole doping level. $\chi_{s}^{\gamma}$
is negative in the whole range of doping explored and becomes more
and more negative on increasing the doping: the holes doped in the
system not only have mainly $p$ character and $E_{2}$ \emph{location},
but they also form plaquette states of singlet type with the $d$
holes - as predicted by ZR conjecture. The actual contribution of
the band $E_{2}$ ($\chi_{s2}^{\gamma}$) is prevalent, as can be
clearly seen in Fig.~\ref{fig_a}. Upon comparing the non-Fermi-weighted
contributions of the bands $E_{2}$ and $E_{3}$ ($\chi_{s\mathbf{2}}^{\gamma}$
and $\chi_{s\mathbf{3}}^{\gamma}$, respectively), we can clearly
establish the singlet nature of the excitations populating the band
$E_{2}$ and the triplet nature of the excitations populating the
band $E_{3}$. As a matter of fact, the bonding component of the $p$
orbital further splits in a singlet and a triplet components (bands
$E_{2}$ and $E_{3}$, respectively) according to the spin coupling
to the $d$ orbital on the plaquette. This is also confirmed by the
further decomposition in momentum space of $\chi_{s\mathbf{2}}^{\gamma}$
and $\chi_{s\mathbf{3}}^{\gamma}$ reported in Fig.~\ref{fig_b}.
It is worth noting the negative effects of increasing $t_{pp}$ on
the net separation between singlets and triplets (see Figs.~\ref{fig_a}
and \ref{fig_b}).

Finally, comparing our results with XAS oxygen K-edge experiments
\cite{Peets_09}, we confirm the impossibility to observe a saturation
of the low-energy spectral weight in the overdoped region in a three-band
model \cite{Wang_10} without taking $ $the oxygen intrasite Coulomb
repulsion into account (see behavior of $n_{2p}^{h}$, the hole occupation
of $p$ character in band $E_{2}$, in Fig.~\ref{fig_c}). On the
other hand, the systematic reduction of the measured oxygen K-edge
intensity assigned to the upper Hubbard band \cite{Peets_09} is well
described in our formulation (see behavior of $n_{1p}^{h}$, the hole
occupation of $p$ character in band $E_{1}$, in Fig.~\ref{fig_c})
as it can be understood in terms of oxygen-hole spectral-weight transfer
between the upper Hubbard band ($E_{1}$) and the ZRS band ($E_{2}$)
driven by the gain in energy at the basis of the mechanism leading
to the ZRS formation.

All these findings highlight once more the strict connection between
the two components (i.e. the extremely high degree of hybridization)
and confirm the soundness of the reduction process to an effective
single-band model in the underdoped regime.

\begin{figure}[t]
\includegraphics[width=8cm]{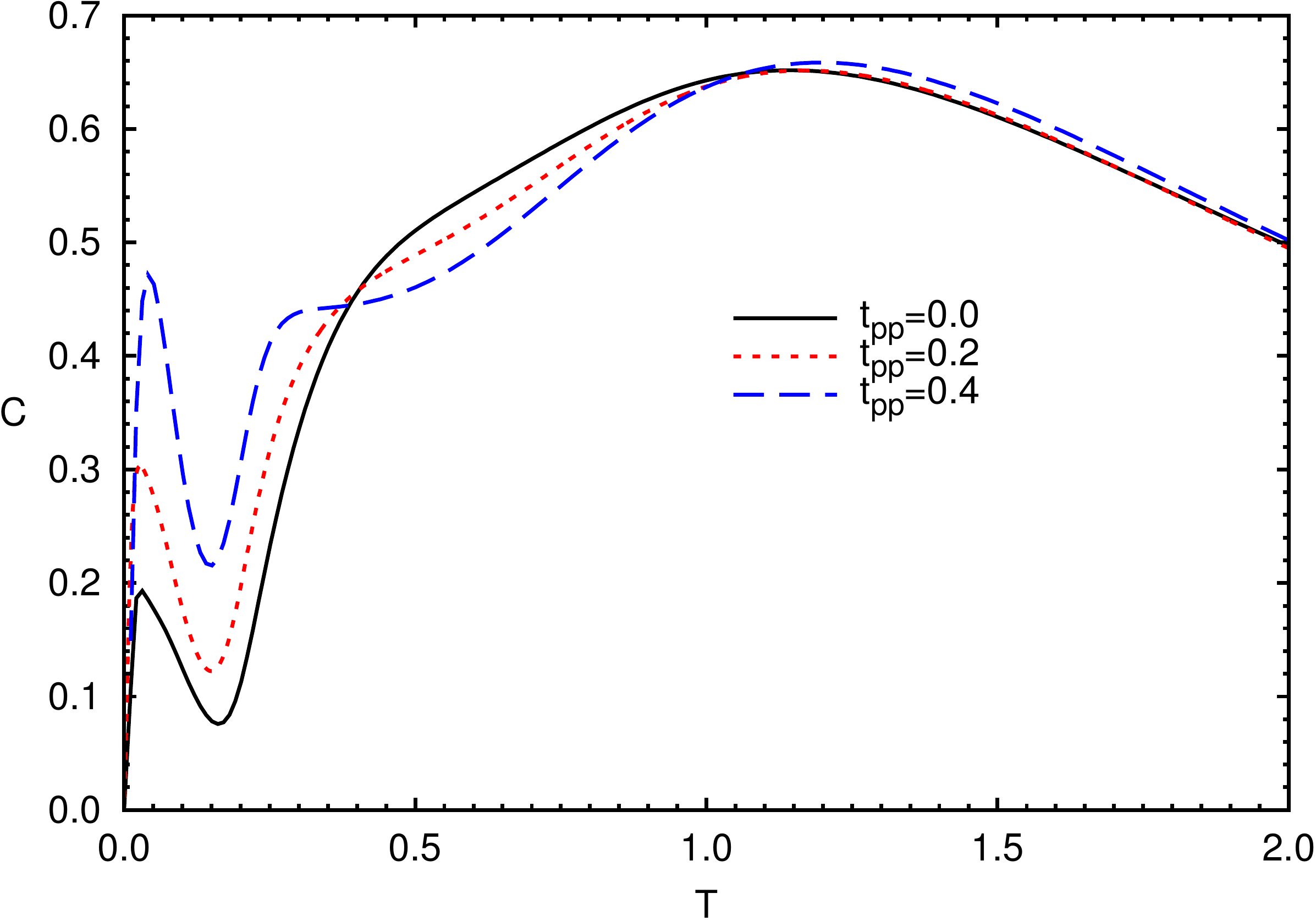}

\includegraphics[width=8cm]{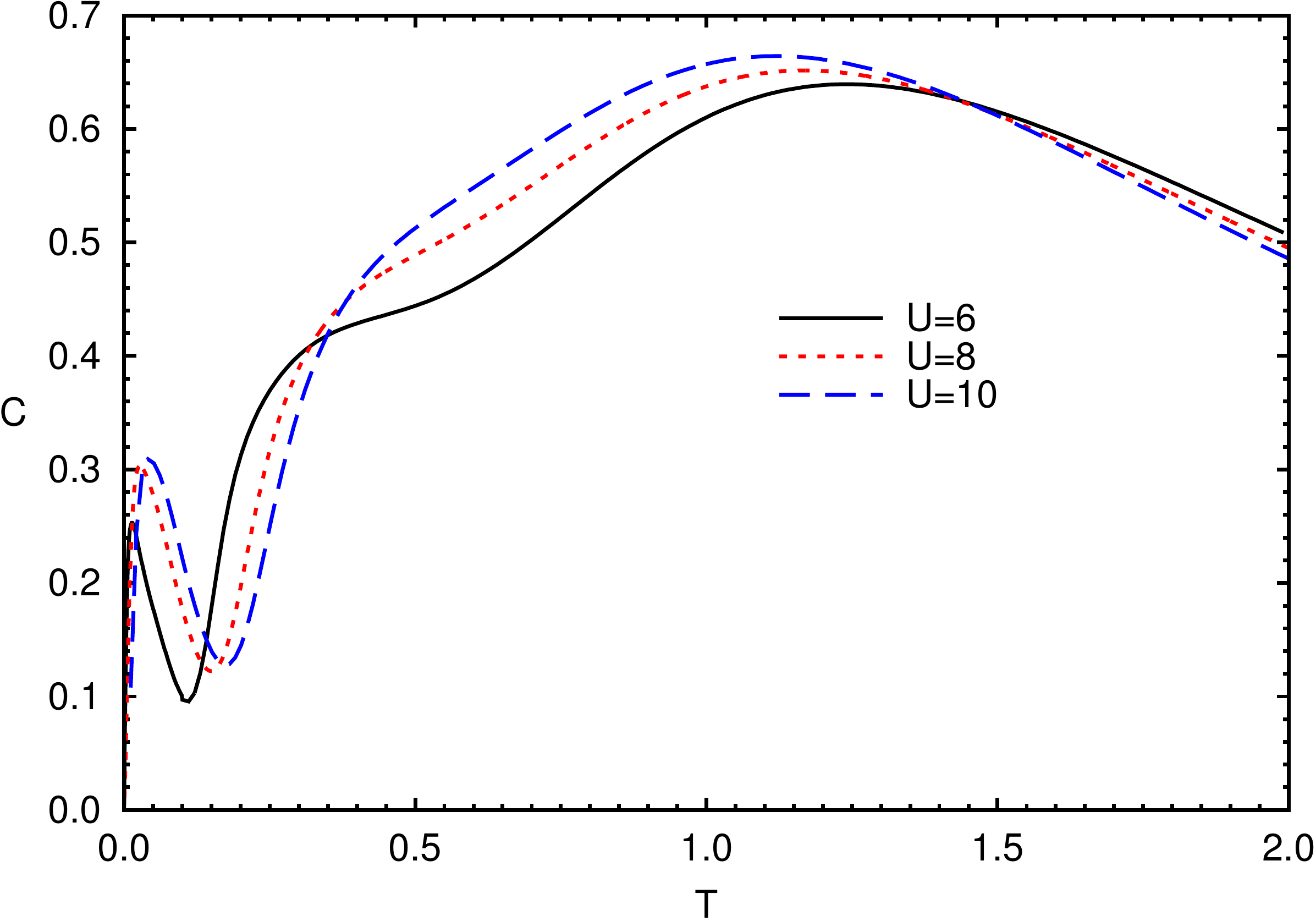} \caption{\label{cs_fig23} (Color online) The specific heat $C$ as a function
of the temperature $T$ for $n_{T}=3$, $\Delta=2$ and: (top panel)
$U=8$ and various values of $t_{pp}$; (bottom panel) $t_{pp}=0.2$
and various values of $U$.}
\end{figure}

\begin{figure}[t]
\includegraphics[width=8cm]{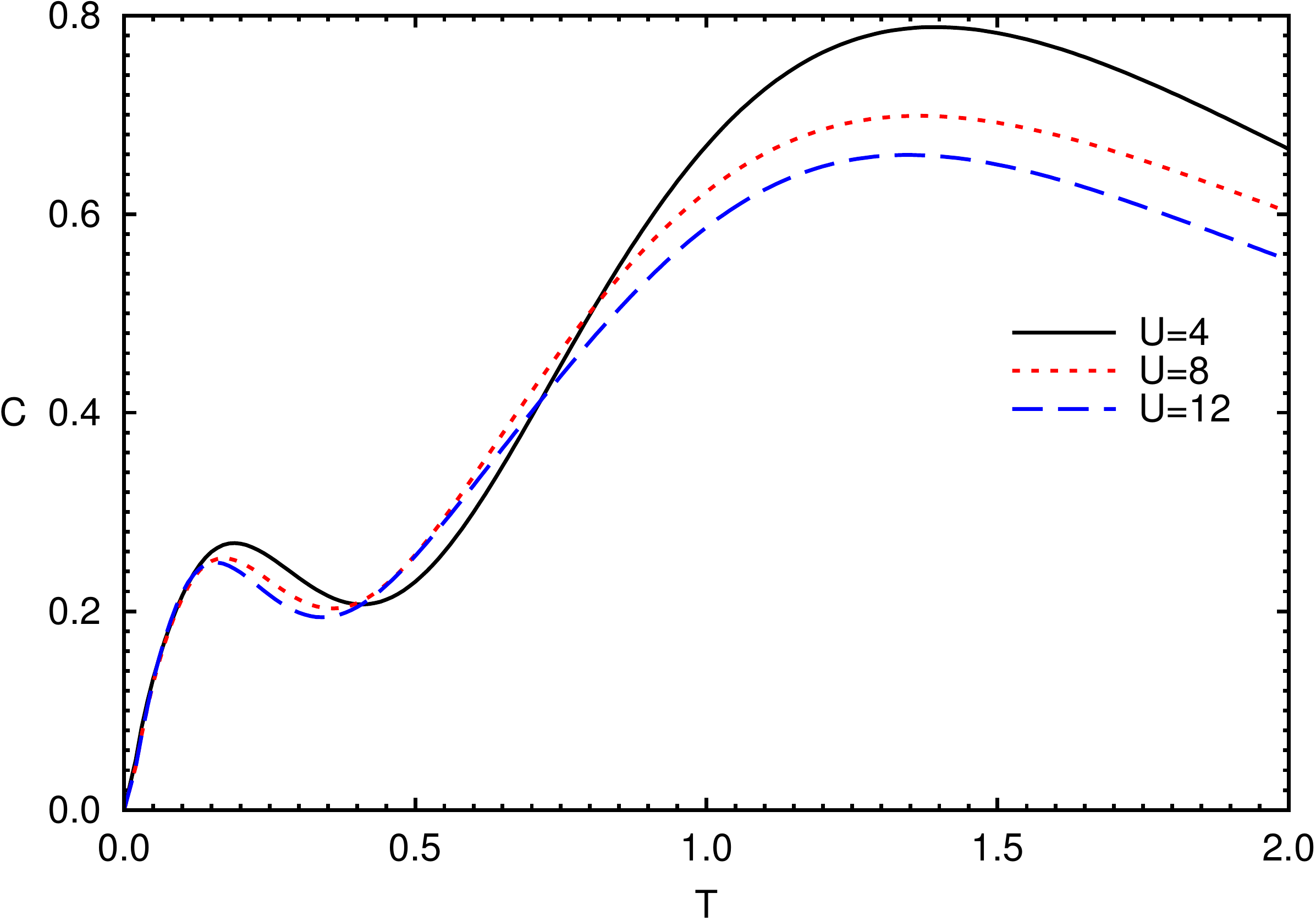}

\includegraphics[width=8cm]{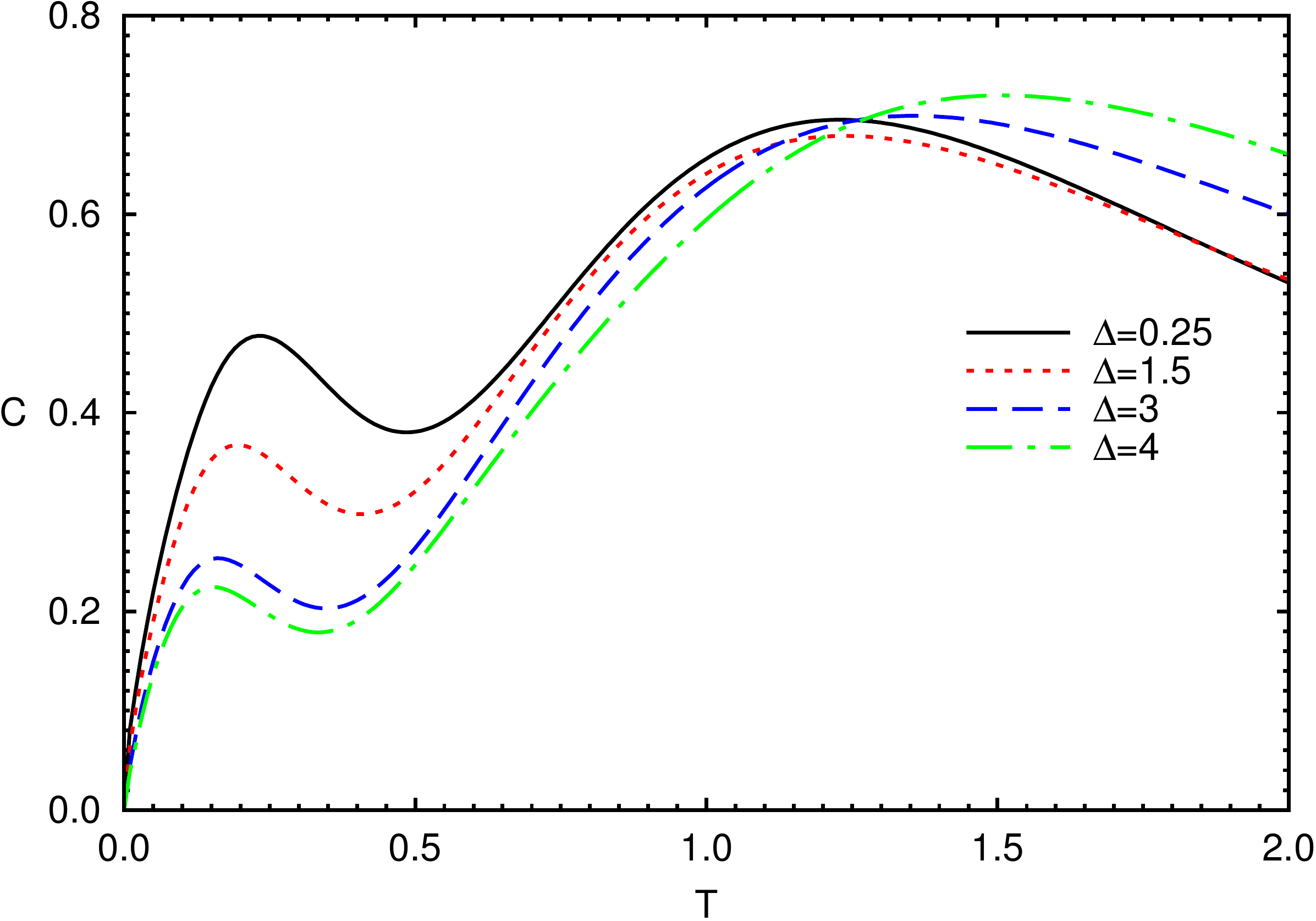} \caption{\label{cs_fig56} (Color online) The specific heat $C$ as a function
of the temperature $T$ for $n_{T}=2.7$, $t_{pp}=0.2$ and: (top
panel) $\Delta=3$ and various values of $U$; (bottom panel) $U=8$
and various values of $\Delta$.}
\end{figure}

\subsection{Thermodynamic properties\label{sec:results:thermo}}

In this section, we report about the thermodynamic properties at finite
temperature of the Emery model. In particular, we focus our attention
on two thermodynamic quantities, namely the specific heat $C$ and
the entropy $S$, and on their temperature dependence on varying the
model parameters. Assuming a different perspective with respect to
that acquired by studying the single-particle properties in the previous
sections, this analysis will allow to collect further pieces of information
on the most relevant energy scales of the system.

\subsubsection{Specific heat\label{spec_heat}}

The specific heat $C$ is defined as $C=dE/dT$, where the internal
energy $E$ can be computed as the thermal average of the Hamiltonian
\eqref{eq1} as 
\begin{equation}
E=\varepsilon_{d}n_{d}+\varepsilon_{p}n_{p}+UD-8t_{pd}(C_{12}^{\gamma}+C_{13}^{\gamma})-4t_{pp}C_{11}^{\lambda}.
\end{equation}

At half filling, $n_{T}=3$, the specific heat presents three peaks
(see Figs.~\ref{cs_fig23}). We wish to remind that a peak in the
specific heat at a certain temperature is usually related to the presence
of an enhancement in the density of states at an energy, with respect
to the chemical potential, about twice larger. The presence of a peak
at the lower temperatures is due to the pinning of the chemical potential
in the proximity either of the very top of the band $E_{2}$ (at $M(\pi,\pi)$)
or of the bottom of the band $E_{1}$ (at $\Gamma(0,0)$) and to the
flatness of the bands at these points that induces very intense peaks
in the density of states (not shown). Such an occurrence enormously
enhances the number of states available at small temperatures and
determines the presence of the corresponding peak in the specific
heat. Consequently, the evident dependence on $t_{pp}$ of both the
position in temperature and the height of this peak (see Fig.~\ref{cs_fig23}
(top panel)) is related to the obvious effects $t_{pp}$ has on the
relative position in energy, with respect to the antidiagonal $X\to Y$,
and flatness of both the top of the band $E_{2}$ and the bottom of
the band $E_{1}$. A similar discussion holds for the dependence on
$U$ as shown in Fig.~\ref{cs_fig23} (bottom panel). At any rate,
this peak is not very fundamental as it can be considered an accident
and much probably would not have any experimental relevance. The second
peak, centered at about $T\approx0.25$ and more evident at large
$t_{pp}$ and small $U$, is connected to the enhancement in the density
of states driven by the van Hove singularity closest to the position
of the chemical potential. Given the tight-binding effective reduction
and the related explanation reported in the previous section, in both
cases (whether the chemical potential lies in $E_{1}$ or $E_{2}$)
this distance in energy is about $0.5$ and varies with $t_{pp}$
and $U$ as shown by the actual position of the peak (see Figs.~\ref{cs_fig23}).
Finally, the third peak, the one centered at about $T\approx1$, is
due to the van Hove singularity of the other band with respect to
the one where the chemical potential lies. In this case, the distance
in energy is mainly due to $\Delta$ and varies very little with $t_{pp}$
while is much more sensitive to $U$, as clearly expected.

Away from half filling, at $n_{T}=2.7$, the specific heat presents
only two peaks (see Figs.~\ref{cs_fig56}). Taking into account that
for this filling the chemical potential lies inside of the band $E_{2}$
and close to its van Hove singularity, the positions in temperature
of these two peaks can be ascribed to the relative positions of the
top of the band $E_{2}$ and to the position of the van Hove singularity
in the band $E_{1}$. It is quite interesting to verify that $U$
does not affect the position and the height of the first peak (see
Fig.~\ref{cs_fig56} (top panel)), as one would expect since the
overall shape and population of the band $E_{2}$ is not very affected
by $U$, while $U$ has a quite visible effect on the height of the
second peak, but not on its position. The average distance between
the two bands, that is, between the two centers of mass, which can
be mainly identified with the positions of the van Hove singularities,
is determined by $\Delta$ and not by $U$ in the charge-transfer
regime, where the model lies for these values of the model parameters.
On the other hand, $U$ definitely reduces the number of available
states in the upper Hubbard sub-band $E_{1}$ and consequently reduces
the height of the second peak. In the metallic regime (for $\Delta<\Delta_{c}$),
$\Delta$ can, instead, enhance quite much the height of the first
peak, although it cannot change its position, as it can influence
the flatness of the top of the band $E_{2}$ in order to accommodate
more and more \emph{moving} particles. On the other hand, $\Delta$
can change the position and, slightly, the height of the second peak
only in the insulating regime (for $\Delta>\Delta_{c}$) as it can
influence the distance of the band $E_{1}$ in order to determine
the charge gap.

According to the above analysis, an experimental measurement of the
specific heat can help determining the values of some of the model
parameters for an effective model.

\begin{figure}[t]
\includegraphics[width=8cm]{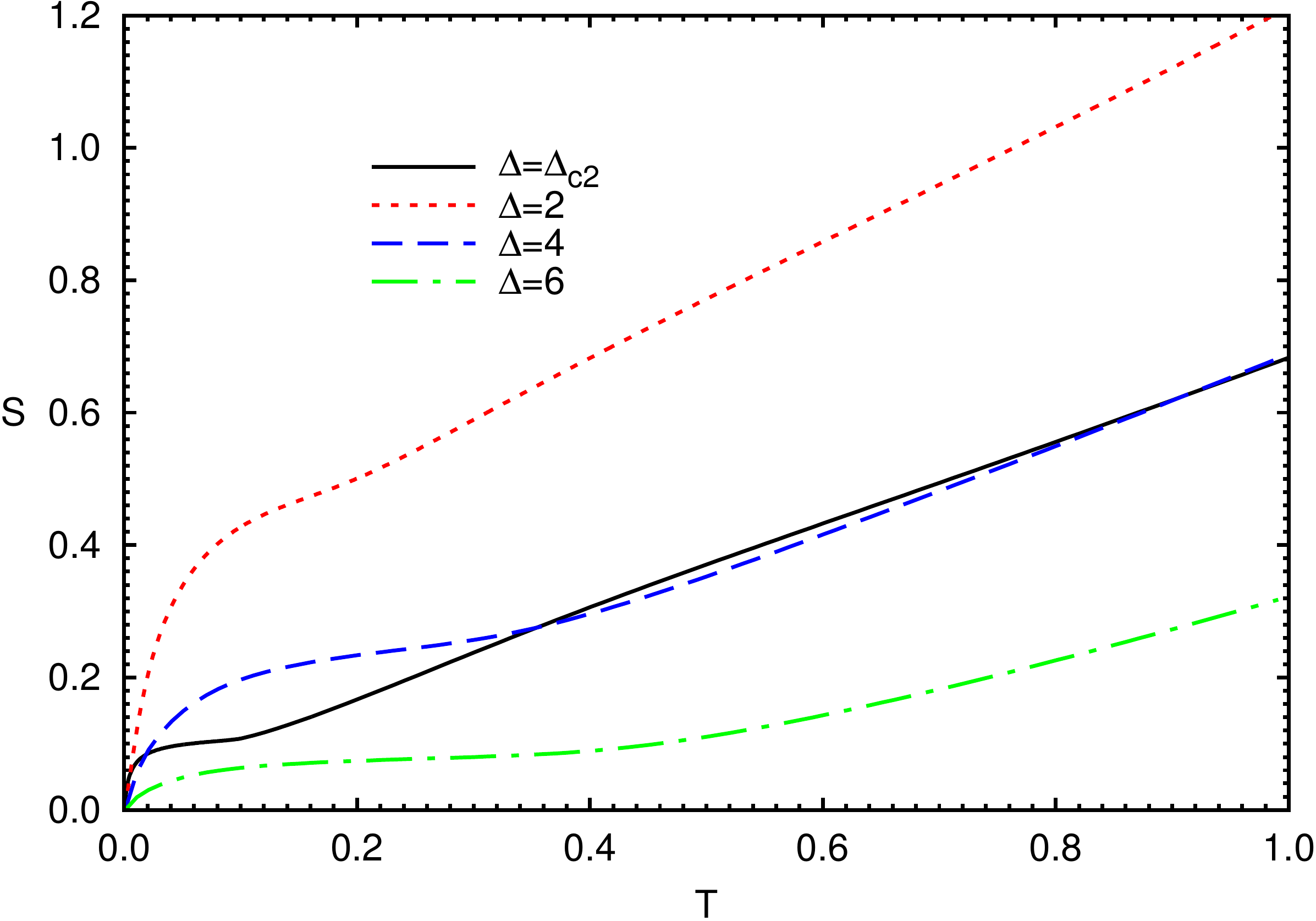} \caption{\label{cs_fig8} (Color online) The entropy $S$ as a function of
the temperature $T$ for $n_{T}=3$, $t_{pp}=0.2$, $U=8$ and various
values of $\Delta$.}
\end{figure}

\subsubsection{Entropy}

The entropy has been computed through the following relation to the
specific heat 
\begin{equation}
S(T)=\int_{0}^{T}\frac{C(T')}{T'}\, dT'.\label{EHM_39}
\end{equation}

In Fig.~\ref{cs_fig8}, we report its behavior as a function of the
temperature at half filling, $n_{T}=3$, in the charge-transfer ($U>\Delta$)
insulating ($\Delta\geq\Delta_{c}$) regime. The great enhancement
at low temperatures is due to the first peak in the specific heat
and, as this latter, can be considered accidental. Definitely more
interesting is the subsequent inflection. It is more or less pronounced
according to the height and capability to resolve the second peak
and is one to one related to the presence of the gap between the two
bands close to the chemical potential and so to the insulating nature
of the system. Accordingly, an experimental measure of the entropy
could give fundamental information about the size of the gap in the
system.

\section{Conclusions\label{sec:conclu}}

In the present manuscript, we have considered the Emery model within
the COM method. With respect to what already done in \cite{Fiorentino_01},
we have introduced a finite direct oxygen hopping $t_{pp}$ and focused
on the single-particle properties in order to discuss the validity
and range of applicability of the ZR construction and scenario. We
first introduced the model and its reduction to the bonding component
only of the oxygen orbital. Then, we checked two possible choices
for the basic field and the related self-consistency equations. Both
choices have been found to give results in very good agreement with
the available numerical ones, being one of the two choices (the one
used in the rest of the manuscript) more numerically stable and so
suitable for systematic and massive use.

Within COM, we observe a metal-insulator transition at half filling
($n_{T}=3$) and our critical line agrees quite well with the one
obtained within DMFT~\cite{Ono_01}. Two regimes can be clearly distinguished:
a Mott-Hubbard regime ($U<\Delta$) and a charge-transfer one ($U>\Delta$).
The ab-initio deduced parameter values relevant for cuprates bring
us into the charge-transfer regime without any adjustment, unlike
many other analytical methods. We also studied the influence of the
finite direct oxygen hopping on the MIT.

Analyzing the single-particle properties of the model, we validated
the ZR scenario as we found the first hole excitations to be of mainly
oxygen type at $(\pi,\pi)$ and the first electronic excitations to
be of mainly copper type at $(0,0)$. Moreover, the reduction to an
effective two-band model resulted definitely feasible as the doping
goes all in one of the four bands, compensates correctly between oxygen
and copper given the very high degree of hybridization found both
in energy and momentum, and the tight-binding effective hoppings results
independent on $U$ and $\Delta$. We also found that the overall
shape of the bands close to the chemical potential is quite similar
to that given by the available ab-initio calculations.

Finally, the analysis of the specific heat and of the entropy allowed
to determine a strict connection between the features of these latter,
in principle experimentally measurable, and the relevant model parameters.
\begin{acknowledgments}
We acknowledge the CINECA award under the ISCRA initiative (project
MP34dTMO), for the availability of high performance computing resources
and support.
\end{acknowledgments}
\appendix

\section{Details of calculations}

\subsection{Basis I\label{app:basisI}}

If we do consider the basis (\ref{eq8}), the normalization matrix
$I(\mathbf{k})$ has the expression 
\begin{equation}
I(\mathbf{k})=\left({\begin{array}{cccc}
1 & 0 & 0 & 0\\
0 & {I_{22}} & 0 & 0\\
0 & 0 & {I_{33}} & 0\\
0 & 0 & 0 & {I_{44}}
\end{array}}\right)\label{mmatr}
\end{equation}
where
\begin{equation}
\begin{split}I_{22} & =1-\frac{n_{d}}{2}\\
I_{33} & =\frac{n_{d}}{2}\\
I_{44} & (\mathbf{k})=3[n_{d}-2D-\chi{\kern1pt }_{s}\alpha(\mathbf{k})]+4a_{s}-\frac{9c^{2}}{I_{22}}-\frac{9b^{2}}{I_{33}}
\end{split}
\end{equation}
with 
\begin{equation}
{\begin{array}{lll}
b & = & \langle p^{\gamma}(i)\eta^{\dag}(i)\rangle\\
c & = & \langle p^{\gamma}(i)\xi^{\dag}(i)\rangle\\
n_{d} & = & \langle n^{d}(i)\rangle
\end{array}}\quad\;{\begin{array}{lll}
D & = & \langle n_{\uparrow}^{d}(i)n_{\downarrow}^{d}(i)\rangle\\
a_{s} & = & \langle p^{\gamma}(i)p^{\dagger\gamma}(i)\sigma_{k}n_{k}^{d}(i)\rangle\\
\chi_{s} & = & \langle n_{k}^{d}(i)n_{k}^{d^{\alpha}}(i)\rangle/3.
\end{array}}
\end{equation}

The matrix $m(\mathbf{k})$ has the expression 
\begin{equation}
m(\mathbf{k})=\left({\begin{array}{cccc}
{m_{11}} & {m_{12}} & {\phantom{-}m_{13}} & 0\\
{m_{12}} & {m_{22}} & {\phantom{-}m_{23}} & {\phantom{-}m_{24}}\\
{m_{13}} & {m_{23}} & {\phantom{-}m_{33}} & {-m_{24}}\\
0 & {m_{24}} & {-m_{24}} & {\phantom{-}m_{44}}
\end{array}}\right).
\end{equation}
and its entries are given by
\begin{equation}
\begin{split}m_{11} & =(\varepsilon_{p}-\mu)+2t_{pp}\lambda(\mathbf{k})\\
m_{12} & =2t_{pd}I_{22}\gamma(\mathbf{k})\\
m_{13} & =2t_{pd}I_{33}\gamma(\mathbf{k})\\
m_{22} & =(\varepsilon_{d}-\mu)I_{22}+2t_{pd}(c-b)\\
m_{23} & =-2t_{pd}(c-b)\\
m_{24} & =2t_{pd}I_{\pi p_{s}}(\mathbf{k})\\
m_{33} & =(\varepsilon_{d}-\mu+U)I_{33}+2t_{pd}(c-b)\\
m_{44} & =(\varepsilon_{p}-\mu)I_{44}(\mathbf{k})+t_{p}I_{\pi p_{s}}(\mathbf{k})\\
 & +2t_{pd}I_{\kappa_{s}p_{s}}(\mathbf{k})+2t_{pp}I_{\lambda_{s}p_{s}}(\mathbf{k})
\end{split}
\label{mmatr2}
\end{equation}
where 
\begin{equation}
\begin{split}I_{\pi p_{s}} & =\frac{3}{2}[n_{d}-\chi_{s}\alpha(\mathbf{k})]+\frac{1}{2}\hat{a}_{s}\\
 & +3(c-b)(bI_{33}^{-1}-cI_{22}^{-1})\\
I_{\kappa_{s}\xi} & =\frac{3}{2}[n_{d}+\hat{a}_{s}-\chi_{s}\alpha(\mathbf{k})]-\frac{3c}{2t_{pd}}(U-\Delta)\\
I_{\kappa_{s}\eta} & =-\frac{3}{2}[n_{d}+\hat{a}_{s}-\chi_{s}\alpha(\mathbf{k})]+\frac{3b}{2t_{pd}}\Delta\\
I_{\lambda_{s}p_{s}} & =4a_{s\lambda}+3cc_{\lambda}I_{22}^{-1}+3bb_{\lambda}I_{33}^{-1}\\
 & +3(n^{d}-2D)-6\chi_{s}\alpha(\mathbf{k})+3\chi_{s\beta}\beta(\mathbf{k})\\
I_{\kappa_{s}p_{s}} & =6[-b-f+d_{s}\alpha(\mathbf{k})]-4b_{s}\\
 & +\frac{9}{2}(bI_{33}^{-1}-cI_{22}^{-1})[n_{d}+\hat{a}_{s}-\chi_{s}\alpha(\mathbf{k})]\\
 & +\frac{9}{2t_{pd}}[c^{2}I_{22}^{-1}(U-\Delta)-b^{2}I_{33}^{-1}\Delta]
\end{split}
\label{mmatr3}
\end{equation}
with 
\begin{equation}
\begin{split}b_{s} & =\langle d^{\alpha}(i)p^{\dagger\gamma}(i)\sigma_{k}n_{k}^{d}(i)\rangle\\
c_{\lambda} & =\langle p^{\gamma\lambda}(i)\xi^{\dag}(i)\rangle\\
b_{\lambda} & =\langle p^{\gamma\lambda}(i)\eta^{\dag}(i)\rangle\\
f & =\langle p^{\gamma}(i)d^{\dag}(i)p^{\gamma}(i)p^{\gamma^{\dag}}(i)\rangle\\
d_{s} & =\frac{1}{3}\langle n_{k}^{d\alpha}(i)\sigma_{k}p^{\gamma}(i)d^{\dag}(i)\rangle\\
\hat{a}_{s} & =-2D+\frac{4}{3}a_{s}\\
a_{s\lambda} & =\langle p^{\gamma\lambda}(i)p_{s}^{\dagger}\rangle\\
\chi_{s\beta} & =\frac{1}{3}\langle n_{k}^{d}(i)n_{k}^{d{\beta}}(i)\rangle.
\end{split}
\end{equation}

By making use of Eqs.~(\ref{mmatr3}), it can be shown that the matrix
element $m_{44}$ can be exactly expressed as $m_{44}=m_{44}^{0}+\alpha(\textbf{k})m_{44}^{\alpha}+\beta(\textbf{k})m_{44}^{\beta}$
with $m_{44}^{0}$, $m_{44}^{\alpha}$ and $m_{44}^{\beta}$ defined
as:

\[
\begin{split}m_{44}^{(0)} & =(\varepsilon_{p}-\mu)I_{44}^{(0)}+9[c^{2}I_{22}^{-1}(U-\Delta)-b^{2}I_{33}^{-1}\Delta]\\
 & -4t_{pd}(3b+3f+2b_{s})\\
 & +6t_{pd}(bI_{33}^{-1}-cI_{22}^{-1})(3n_{d}+\hat{a}_{s})\\
 & +18t_{pd}(c-b)(bI_{33}^{-1}-cI_{22}^{-1})^{2}\\
 & +2t_{pp}[4a_{s\lambda}+3cc_{\lambda}I_{22}^{-1}+3bb_{\lambda}I_{33}^{-1}+3(n_{d}-2D)]
\end{split}
\]
\[
\begin{split}m_{44}^{(\alpha)} & =(\varepsilon_{p}-\mu)I_{44}^{(1)}\\
 & +6t_{pd}[2d_{s}-3(bI_{33}^{-1}-cI_{22}^{-1})\chi_{s}]-12t_{pp}\chi_{s}\\
m_{44}^{(\beta)} & =6t_{pp}\chi_{sb},
\end{split}
\]
where we have introduced the quantities $I_{44}^{(0)}$ and $I_{44}^{(1)}$,
which read as follows 
\begin{equation}
\begin{split}I_{44}^{(0)} & =3n_{d}+\hat{a}_{s}-9c^{2}I_{22}^{-1}-9b^{2}I_{33}^{-1}\\
I_{44}^{(1)} & =-3\chi_{s}.
\end{split}
\end{equation}

\subsection{Basis II\label{app:basisII}}

If we do consider the basis (\ref{eq80}), the normalization matrix
$I(\mathbf{k})$ has the expression given in~(\ref{mmatr}) but with
$I_{44}(\mathbf{k})$ given by 
\[
I_{44}(\mathbf{k})=a-I_{33}^{2}-I_{22}^{-1}g^{2}-I_{33}^{-1}g^{2}+f_{s}+(I_{33}^{2}-p)\alpha(\mathbf{k})
\]
where 
\[
\begin{split}g & =\langle p^{\gamma}(i)\xi^{\dag}(i)\rangle-\langle p^{\gamma}(i)\eta^{\dag}(i)\rangle\\
a & =1-\langle p^{\gamma}(i)p^{\dag}(i)\rangle-2\langle\pi(i)p^{\gamma\dag}(i)\rangle\\
f_{s} & =2\langle\sigma_{k}n_{k}^{d}(i)p^{\gamma}(i)p^{\gamma\dag}(i)\rangle\\
p & =\frac{1}{4}\langle n_{\mu}^{d}(i)n_{\mu}^{d^{\alpha}}(i)\rangle-\langle d_{\uparrow}(i)d_{\downarrow}(i)[d_{\downarrow}^{\dag}(i)d_{\uparrow}^{\dag}(i)]^{\alpha}\rangle
\end{split}
\]
The matrix $m(\textbf{k})$ has the same expression given in~(\ref{mmatr2})
and~(\ref{mmatr3}) but with $m_{44}(\textbf{k})$ given by 
\[
m_{44}(\textbf{k})=m_{44}^{(0)}+m_{44}^{(\alpha)}\alpha(\textbf{k})+m_{44}^{(\beta)}\beta(\textbf{k})
\]
where 
\[
m_{44}^{(\beta)}=\frac{1}{4}\langle n_{\mu}^{d}(i)n_{\mu}^{d{\beta}}(i)\rangle-\langle d_{\uparrow}(i)d_{\downarrow}(i)[d_{\downarrow}^{\dag}(i)d_{\uparrow}^{\dag}(i)]^{\beta}\rangle
\]
The quantities $m_{44}^{(0)}$ and $m_{44}^{(\alpha)}$ are calculated
by imposing the algebra constraints embedding the Pauli principle.
\bibliographystyle{apsrev}
\bibliography{pd}
 
\end{document}